\documentclass[prmaterials,aps,preprint,showkeys,longbibliography]{revtex4-2}

\usepackage{epsfig,amsmath,amssymb,txfonts,hyperref,xr,physics}
\usepackage{xcolor}
\usepackage{multirow}

\newcommand{\Fig}[1]{Fig.~\ref{fig:#1}}

\newcommand{\Tab}[1]{Table~\ref{tab:#1}}

\newcommand{\Eqn}[1]{Eqn.~\ref{eqn:#1}}

\newlength{\wholefigwidth}
\newlength{\halfwholefigwidth}
\newlength{\figwidth}
\date{\today}

\begin{document}

\title{Core Energies in Isolated Edge and Mixed Dislocations in BCC Fe from First-principles Energy Density Method}

\author{Yang Dan}
\email{yangdan2@illinois.edu}
\author{Dallas R. Trinkle}
\email{dtrinkle@illinois.edu}
\affiliation{Department of Materials Science and Engineering, University of Illinois, Urbana-Champaign, Urbana, Illinois 61801, USA}

\begin{abstract}
    We use first-principles spin-polarized energy density method (EDM) to calculate the atomic energies in isolated $a_0[100](010)$ edge, $a_0[100](011)$ edge, $\frac{a_0}{2}[\bar{1}\bar{1}1](1\bar{1}0)$ edge and $\frac{a_0}{2}[111](1\bar{1}0)$ $71^\circ$ mixed dislocations in body-centered cubic (BCC) Fe. The distribution of atomic energies shows the energetic effects of slip, including non-linear displacements in and near the core, and the elastic field further away.
    The EDM atomic energies agree well with anisotropic elasticity predictions in the elastic region, while they deviate in the core region due to failure of linear elasticity, and the energy deviation quantifies the core widths. Dislocation line energies are obtained by partial sums of the atomic energies within distance $r$ of the dislocation center; the core energy is extracted from the large $r$ behavior. We compare our results with an EAM and a GAP potential, showing that while both potentials produce core structures similar to DFT predictions, the GAP potential has a closer match with DFT and EDM in core energies.
\end{abstract}

\maketitle

Plastic deformation is a critical mechanical behavior of structural metals, and is governed by dislocation motion or slip. In many steel alloys, iron is present in the body-centered cubic (BCC) phase, and so dislocation behavior in that phase is of utmost importance for understanding deformation\cite{Berns2008}. While deformation at low temperature is controlled by $\frac12 a_0 \langle 111 \rangle$-type screw dislocations due to their low mobility due to high Peierls barriers from their nonplanar core structures\cite{Christian1983,Cai2004}, other types of dislocations also control slip. For example, $\frac12 a_0 \langle 111 \rangle$-type edge dislocations have different Peierls stresses and are typically more mobile\cite{Nabarro2002}. As they move in the material, they can react with other $\frac12 a_0 \langle 111 \rangle$-type dislocations via the favorable reaction\cite{Puschl1985,Schoeck2010}
\begin{equation}
  \frac{a_0}{2}[111] + \frac{a_0}{2}[1\bar{1}\bar{1}] \rightarrow a_0[100],
  \label{eqn:reaction}
\end{equation}
forming stable binary junctions with the $a_0 \langle 100 \rangle$ Burgers vector, which are less mobile themselves or are possible to further react to form ternary junctions, leading to dislocation locks that impede deformation and contribute to work hardening. Edge dislocations participate in dynamical processes that influence the formation of dislocation loops\cite{Bonny2016}, and are known to interact with dislocation loops, which impacts the evolution of microstructures under cascade damage conditions\cite{Osetsky2000}. The $70.5^\circ$ mixed (M111) dislocation has also been identified with a high Peierls barrier despite its planar core structure\cite{Kang2012,Romaner2021}.

The behaviors of dislocations during motion and reactions can be understood and predicted in part from their energies. For example, the stable core structure comes with the minimal energy; junction formation and dislocation reaction occur in the direction that reduces the configuration energy; the mobility of a dislocation is controlled by the Peierls barrier, which is the energy barrier of a dislocation in their lowest-energy gliding path. The core energy is an important component in the dislocation energy and is especially important for the core configurations and behaviors, but it cannot be derived from linear elasticity theories due to the large lattice distortions.

The line energy (energy per unit length) of an isolated dislocation in an infinite medium is a combination of the elastic energy and the core energy. Within a cylindrical region of radius $R$ around the dislocation line, the line energy $E_\text{disloc}$ can be written as the sum of the elastic component and the core component
\begin{equation}
    E_\text{disloc}(R) = \frac{Kb^2}{4\pi} \ln \frac{R}{r_c} + E_\text{core},
    \label{eqn:E_disloc}
\end{equation}
where $K$ is the energy prefactor that can be determined from elastic constants based on anisotropic elasticity theory\cite{Hirth1982}, $b$ is the Burgers vector, and $r_c$ is the size of the core. The elastic energy diverges due to the elastic discontinuity at the core, so a cut-off $r_c$ is introduced, and energy contributions from atoms in the core with $R<r_c$ are included in $E_\text{core}$. There are no analytical solutions for $E_\text{core}$, and if $E_\text{core}$ is to be estimated from the dislocation total energy, proper ways to separate the elastic contributions from the total energy are necessary.

Unlike point or planar defects, extracting the dislocation energy from the total energy difference of a supercell with and without defects is generally difficult and complex for dislocations. This is due to the inclusion of domain boundaries (screw dislocations) or free surfaces (mixed and edge dislocations) that are not included in \Eqn{E_disloc}. To remove such surface contributions, dislocation multipoles (dipoles, quadrupoles, etc.) that cancel out the long-ranged strain field are arranged in the same supercell. The combined energy of the multipole is well captured by the supercell total energy; however, several forms of elastic energy contributions need to be accurately modeled and excluded from the multipole energy to obtain the core energy. These contributions include the elastic energy in areas away from the multipole, the interaction energy within the multipole, the image interactions due to periodic boundary conditions, and the energy of an elastic ``core field'' for fast convergence in the core energy within reasonable supercell sizes\cite{Clouet2009Ecore,Clouet2011Ecore}. Previous works have followed such routines to study dislocation core energies in BCC Fe. Clouet \textit{et al.} introduced a short-range elastic core field in addition to the Volterra elastic field in a geometry containing dislocation dipoles, and extract the core energies by subtracting elastic interaction energies between the dipoles and their images along with self-elastic energies of the core fields from the first-principles system total energy, which is applied to $\langle 111 \rangle$ screw dislocations and is able to calculate other dislocation types\cite{Clouet2009Ecore,Clouet2011Ecore}. The total energy difference method is also applicable to modelings with interatomic potentials. For example, Bertin \textit{et al.} extracted core energies from dislocations with different character angles with atomic interactions modeled by different interatomic potentials\cite{Bertin2021}.

A possible way to overcome the difficulty and complexity of the total energy difference method is to instead consider the energy of atoms inside of a cylindrical region of radius $R$ around the dislocation line. For classical interatomic potentials, the per-atom potential energy can be used as such local energies to sum up for the core energy, but transferability issues lead to disagreement among different interatomic potentials trained with different methods and datasets; in fact, many interatomic potentials fail to reproduce the core structures modeled by DFT\cite{Fellinger2018}. For first-principles methods like DFT, local energies have non-unique definitions and any definition that sums up to be the Kohn-Sham total energy is considered valid, due to gauge dependence. The energy density method (EDM) provides a possible insight for partitioning the Kohn-Sham total energy into well-defined atomic energies\cite{Yu2011EDM,Dan2022}. The first-principles energy densities are derived from the formalism of DFT, and are integrated over atomic volumes that are partitioned to individual atoms where the gauge-dependent terms integrate to zero. This approach ensures the well-defined and unique local energies for the atoms in a supercell. Isolated dislocations avoid extra calculations to assess energy contributions from elastic interactions as in the case of dislocation dipoles or multipoles, and therefore it is more straightforward; however, the spurious energy contribution from the domain boundary or free surface cannot be separated from the core energy if only the total energy is used. With EDM, it is possible to obtain energies in local portions of the geometry only for an arbitrary type of dislocation modeled by DFT, thus it easily excludes energies away from the dislocation core\cite{Dan2022}.

In this work, we study the core energies of $a_0[100](010)$ edge, $a_0[100](011)$ edge, $\frac{a_0}{2}[\bar{1}\bar{1}1](1\bar{1}0)$ edge and $71^\circ$ $\frac{a_0}{2}[111](1\bar{1}0)$ mixed dislocations using EDM. These three edge dislocations are the most commonly observed in experiments in BCC Fe\cite{Clouet2008,Fellinger2018}. A major set of slip planes in BCC Fe are $\{ 110 \}$ planes\cite{Franciosi1983}, which involve both edge and mixed dislocations depending on the direction of the Burgers vector, and we study the $71^\circ$ $\frac{a_0}{2}[111](1\bar{1}0)$ mixed dislocation as a representative for the mixed dislocations. In contrast to most previous studies of dislocation core energies, isolated dislocation geometries are used. Our result will serve as a comparison to previous methods involving dipoles or multipoles, provide benchmarking energy data to future classical potential developments, and showcase as an example for core energy calculation with EDM for general types of dislocations, which benefits other dislocation simulations, such as discrete dislocation dynamics (DDD) simulations. We first show the atomic energy distributions calculated from EDM in the dislocation geometries, with a focus on such energy distributions in the core region and its connection with the core structure. Then we compare atomic energies calculated by EDM and by anisotropic elasticity and show their consistency in the core region; and we further show that their difference helps to derive core widths that reflect geometric features of the cores. Finally, we derive line energies from atomic energies, from which we extrapolate core energies. We compare our results with available data, and discuss the favorability of the reaction in \Eqn{reaction}.

\begin{figure*}[htbp]
    \makebox[\textwidth][c]{%
        \includegraphics[width=\figwidth]{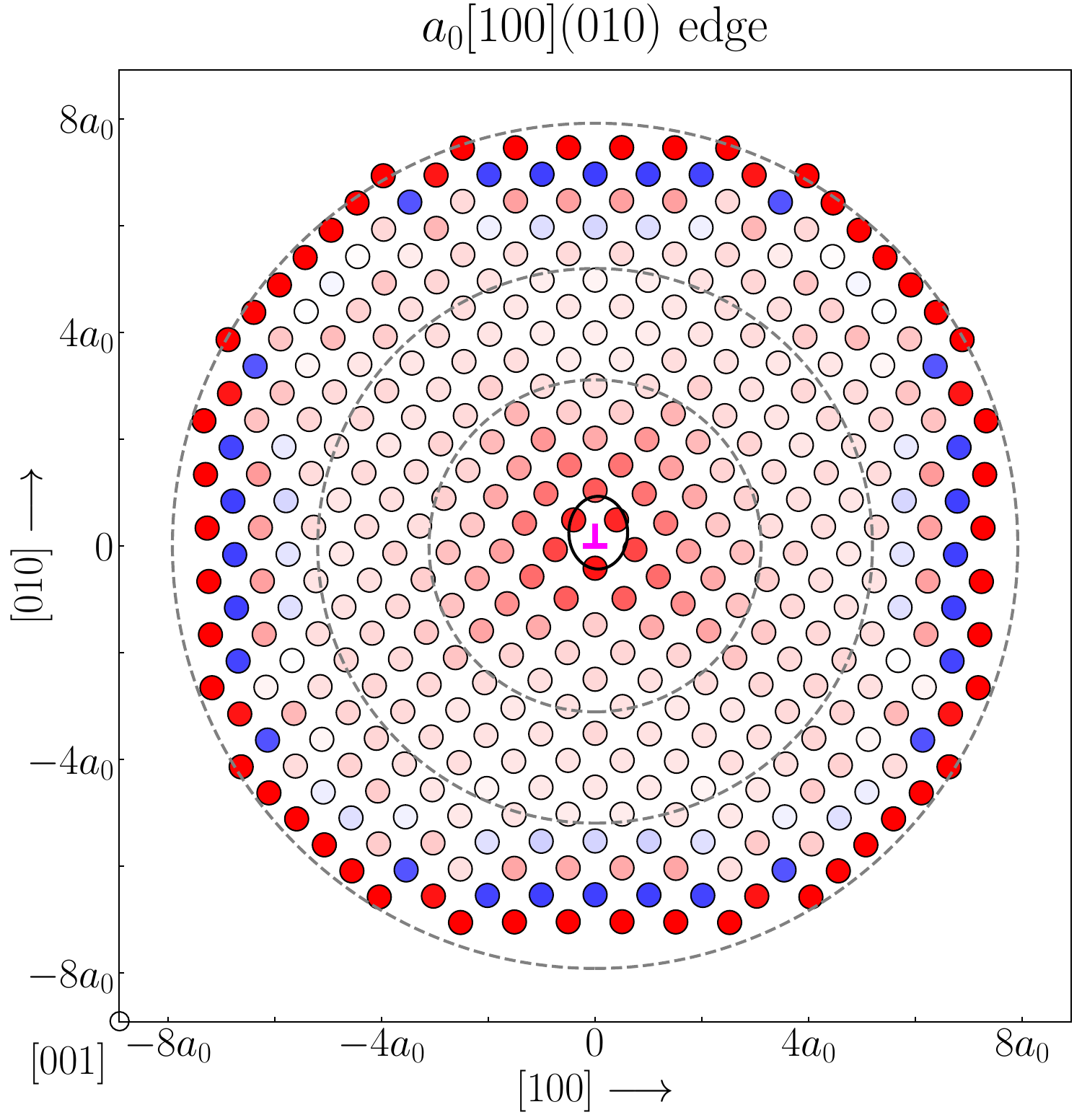}
        \includegraphics[width=\figwidth]{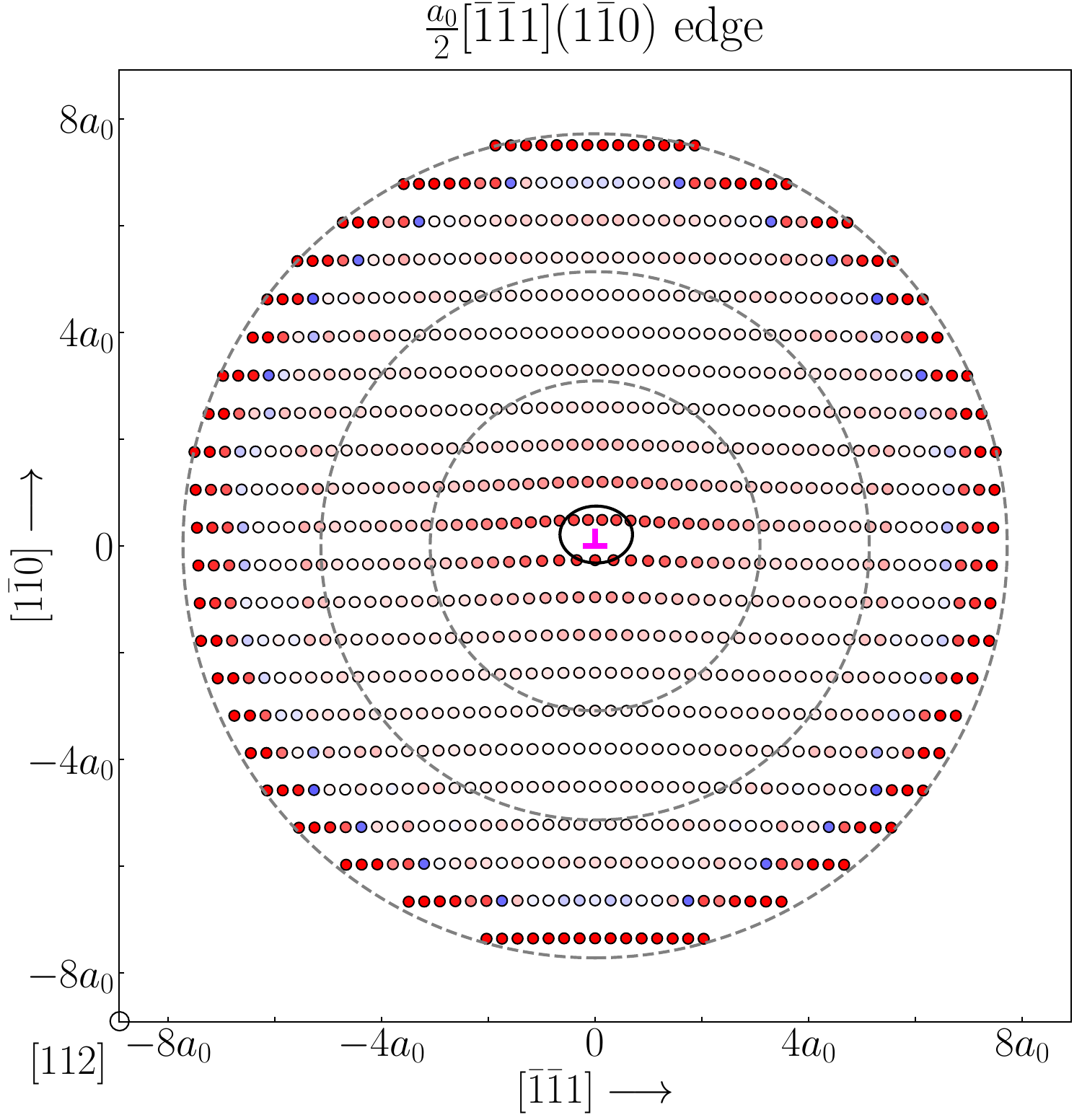}
    }
    \makebox[\textwidth][c]{%
        \includegraphics[width=\figwidth]{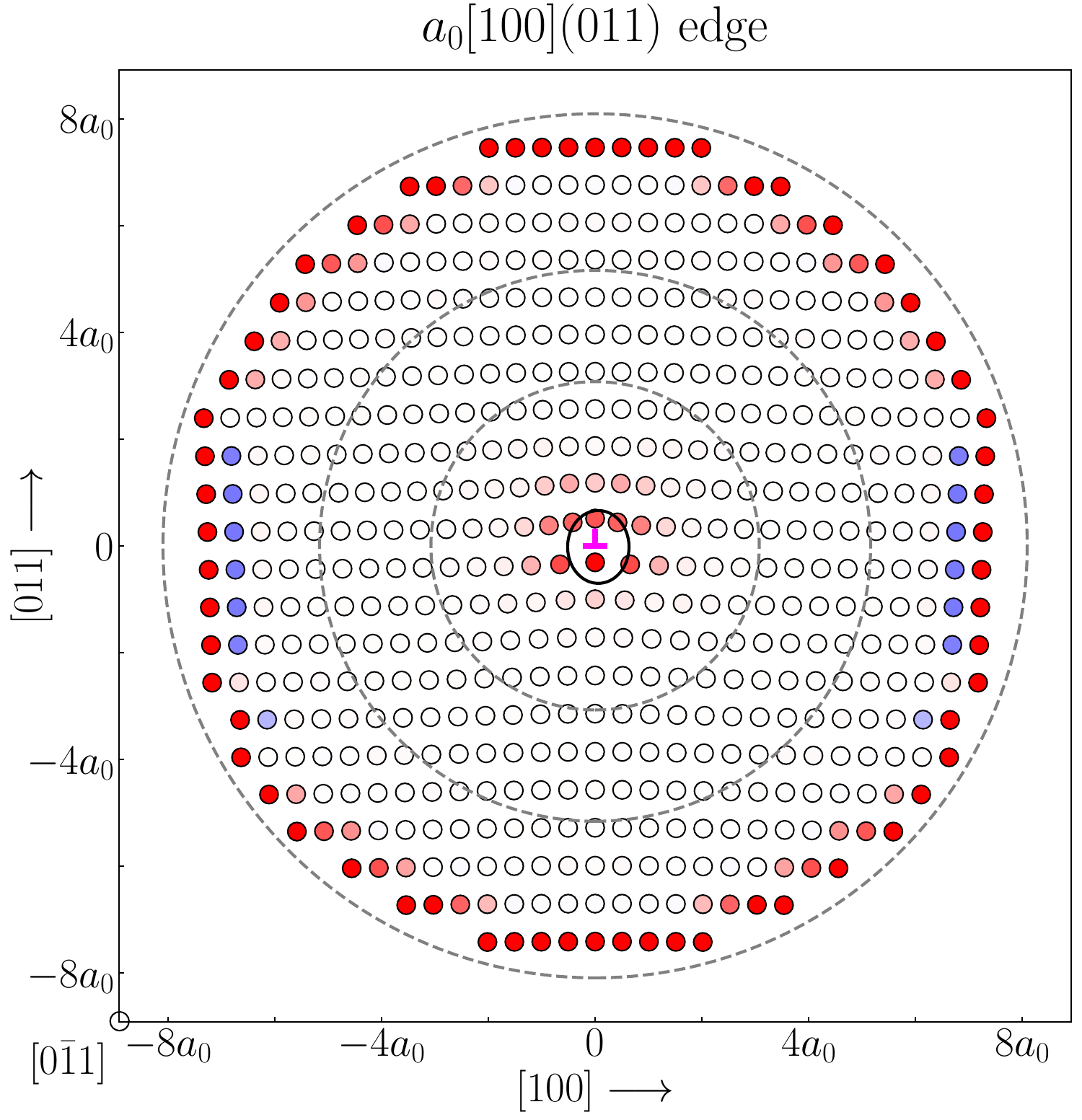}
        \includegraphics[width=\figwidth]{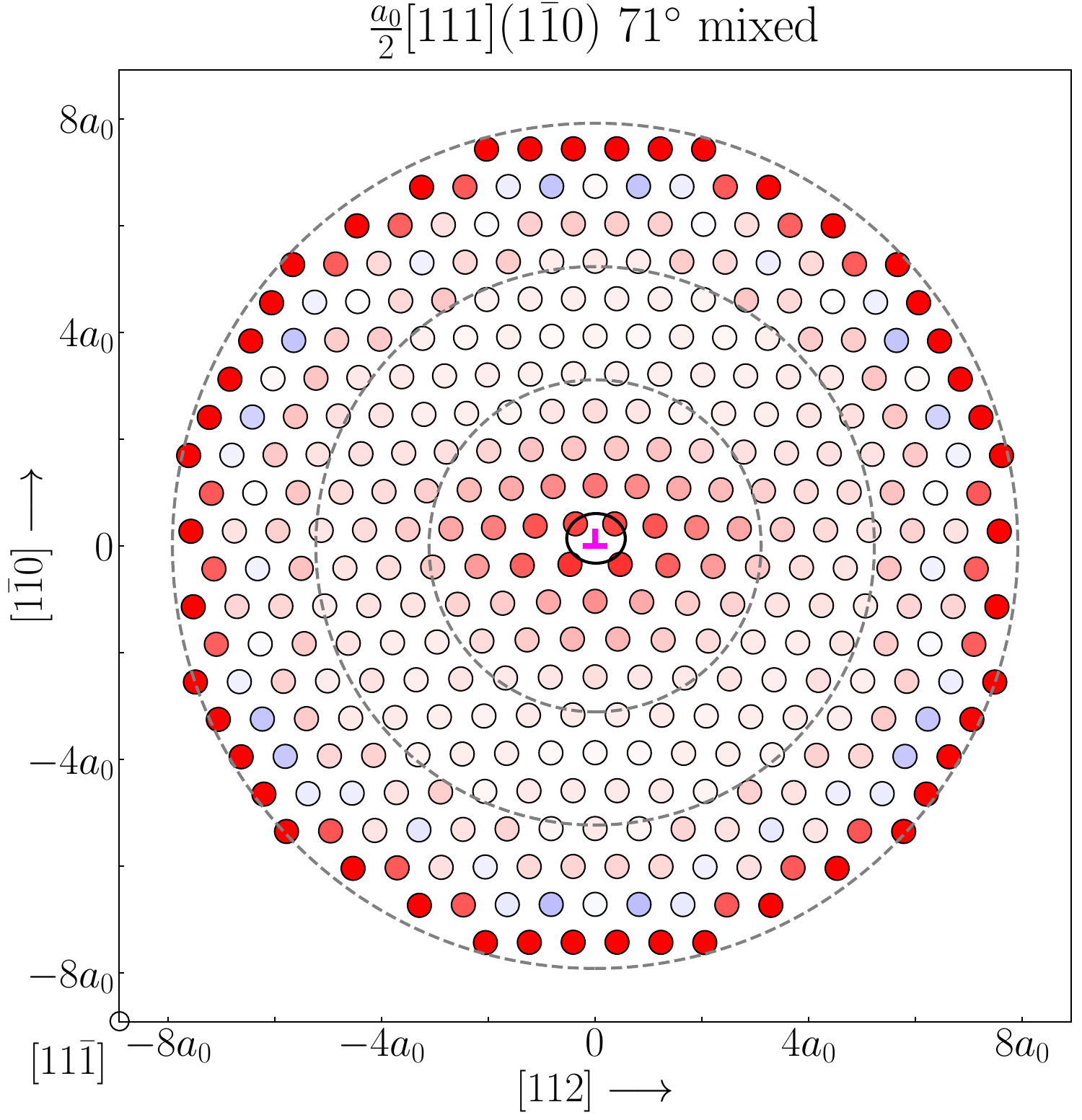}
    }
    \makebox[\textwidth][c]{%
        \qquad
        \includegraphics[width=\figwidth]{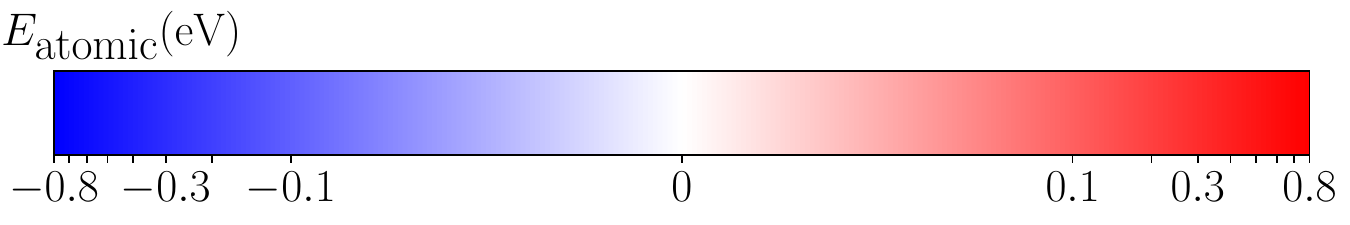}
    }
    \caption{Supercells of geometries of $a_0[100](010)$ edge, $\frac{a_0}{2}[\bar{1} \bar{1} 1](1\bar{1}0)$ edge, $a_0[100](011)$ edge and $\frac{a_0}{2}[111](1\bar{1}0)$ $71^\circ$ mixed dislocations in BCC Fe, which are used for EDM calculations, with EDM energies shown with colors on each atom. The geometries are made by Fellinger \textit{et al.}\cite{Fellinger2018}. The lattice parameter is $a_0 = 2.832 \text{\AA}$ and all supercells are 50.46 \AA\ $\times$ 50.46 \AA\ in dimensions perpendicular to the threading direction that points out of the page, along which the supercells are periodic. The black ``$\perp$''  signs mark the center of each dislocation's elastic displacement field. The gray dashed circles show the region boundaries in FBC, which are centered at the dislocation centers. The black ellipses mark the range of the core widths determined by EDM energy data, which are centered at average positions defined by \Eqn{avgpos}, and their the horizontal and vertical diameters equal the core widths defined by \Eqn{corewidths}. The core widths numbers are listed in \Tab{corewidths}.}
    \label{fig:geometry-edm}
\end{figure*}

Atomic energies are calculated by applying EDM to the isolated $a_0[100](010)$ edge, $a_0[100](011)$ edge, $\frac{a_0}{2}[\bar{1}\bar{1}1](1\bar{1}0)$ edge and $\frac{a_0}{2}[111](1\bar{1}0)$ mixed dislocations, whose geometries were relaxed by Fellinger \textit{et al.}\cite{Fellinger2018} using flexible boundary conditions (FBC)\cite{Sinclair1978FBC,Woodward2005FBC,TrinkleLGF2008,Tan2016}. The initial geometries of dislocations are created by displacing atoms according to the solution of Volterra's equation based on anisotropic elasticity\cite{Bacon1980}. The atoms are then relaxed using a combined approach of DFT and lattice Green function (LGF) so that the atom displacements in the core are coupled to the harmonic bulk in the far field. In this process, as shown in \Fig{geometry-edm}, the geometry of each dislocation is divided into three regions based on the distance to the dislocation center: region I in the vicinity of the core center contains all the atoms inside the core with large displacements, region II away from the core center which surrounds region I and contains atoms with smaller displacements, and region III contains free surfaces. A vacuum region is outside region III, and the thickness of region III is chosen to be larger than the range of atomic interactions so regions I and II are isolated from any effects caused by the vacuum. All the supercells have the same size of $50.46$ \AA\  $\times$ $50.46$ \AA\ for dimensions in the page, and have threading directions perpendicular to the page. Along the threading direction, the supercells are periodic. The relaxations are done iteratively with DFT forces from regions I and II; region I is updated with a conjugate gradient step, while the LGF provides displacements on all atoms from the forces in region II. This continues until forces converge in regions I and II below 5meV/\AA. Then EDM calculations are performed using the relaxed geometries for the atomic energies. The EDM method is implemented into the software package Vienna Ab Initio Simulation Package (VASP)\cite{Kresse1996VASPa,Kresse1996VASPb} (version 5.4.4). The ingredients of EDM (energy densities, charge densities and potentials) are computed after the charge densities self-consistently converge in the electronic optimization; the energy densities are then integrated for atomic energies over volumes partitioned to each atom based on analysis of the charge densities and potentials\cite{Yu2011EDM}. The EDM atomic energies are referenced by the energies of Fe atoms in BCC bulk. The parameters for EDM calculations follow those used in the DFT relaxations for consistency in these series of calculations, except that the real-space grids are created with a uniform spacing of $0.075$ \AA\ in each dimension, which is twice as dense as used in the DFT relaxations, to ensure that the gauge-dependent errors\cite{Yu2011EDM} in EDM atomic energies fall below 1 meV/atom. A plane-wave basis is used with cut-off energy 400 eV to converge the total energy below 1 meV/atom. The energy tolerance for self-consistency calculations is $10^{-8}$ eV. Spin polarization is enabled in all calculations for accurate description of the ferromagnetism in BCC Fe. The projector augmented-wave (PAW) method\cite{Blochl1994PAW,Kresse1999PAW,Hobbs2000PAW} with 8 valence electrons ($[\text{Ar}]3d^74s^1$) is used, with the Perdew-Burke-Ernzerhof (PBE) functional\cite{Perdew1996PBE} for the exchange-correlation energy. Methfessel-Paxton smearing\cite{Methfessel1989} with order 1 and an energy width of 0.25 eV is applied. The k-point mesh grids are gamma-centered with uniform spacing of approximately 0.125 \AA$^{-1}$ in each dimension in the k-space. The smearing and k-point mesh are chosen so that the density of states (DOS) of bulk BCC Fe near the Fermi level well agrees the DOS calculated by the linear tetrahedron method with Bl\"{o}chl corrections\cite{Blochl1994Tetra} and with a dense k-point mesh.

We also compare our DFT and EDM calculations with two interatomic potentials: the EAM potential developed by Mendelev \textit{et al.}\cite{Mendelev2003}, which has the best overall performance in reproducing the DFT dislocation core structures among several other potentials\cite{Fellinger2018}, and the GAP potential developed by Zhang \textit{et al.}\cite{Zhang2024} that has been trained and compared with the same DFT dislocation geometries used in this work. Cylindrical slabs of radius 300 \AA\ are made for each geometry, where the radius is chosen to be large enough to avoid boundary effects on the dislocation cores in the following steps. The dislocation is introduced at the center of the slab by displacing atoms according to the solution of Volterra's equation based on anisotropic elasticity\cite{Bacon1980}, with the threading direction along the axis of the slab. The elastic constants used for anisotropic elasticity are $C_{11} = 243.4$ GPa, $C_{12} = 145.0$ GPa and $C_{44} = 116.0$ GPa for the EAM potential, and $C_{11} = 280.5$ GPa, $C_{12} = 155.8$ GPa and $C_{44} = 101.1$ GPa for the GAP potential, which are determined by fitting the stress response to finite deformations, and are close to the reported values available\cite{Mendelev2003}. We fix the atoms furthest from the dislocation core, at a distance of 15 \AA\ from the outside edge; this exceeds the cut-off distances for interatomic interactions for both potentials. The inner atoms are relaxed first using conjugate gradient (CG) algorithm for a maximum of 300 iterations which converges the forces to $10^{-3}$ eV/\AA; then the FIRE algorithm\cite{Bitzek2006FIRE} is applied for 150 iterations in each dislocation geometry which converges the forces to $10^{-4}$ eV/\AA, followed by applying CG algorithm again for a maximum of 1000 iterations which converges the forces to $10^{-5}$ eV/\AA. The atomic energies are calculated as the per-atom potential energies in the fully relaxed dislocation geometries, and are referenced to the atomic energy in the BCC calculated using the same potential.

\Fig{geometry-edm} shows that for each dislocation, EDM atomic energies are high near the core center in region I and near the free surface in region III, but are small in region II, where the regions for FBC are marked by gray dashed circles. These geometries are shown in the $m-n-t$ coordinate system; the horizontal $m$-axis aligns with the direction of the Volterra cut (the Burgers vector for edge dislocations, and the edge component for the mixed dislocation), the $t$-axis is the dislocation threading direction out of the page, and the slip plane normal $n$ is perpendicular to both. Zero atomic energy is defined as the energy of the Fe atom in the BCC bulk. Atoms near the core center in region I show large positive energies in general, indicating that the lattice distortions from the core significantly increase the local energy. Moving away from the core center, the atomic energies decay to smaller values in region II, which is the result of smaller strain and stress further from the core. Large positive and negative energies are observed in the outermost layers of atoms near the vacuum in region III, which originate from the artificial free surfaces and are spurious. However, these large atomic energies disappear when moving inward towards the boundary between region III and region II, showing that region III is thick enough to isolate regions I and II from the effects of the free surface, which will be discussed in later paragraphs with more details. 

\begin{figure*}[htbp]
    \makebox[\textwidth][c]{%
        \includegraphics[width=\figwidth]{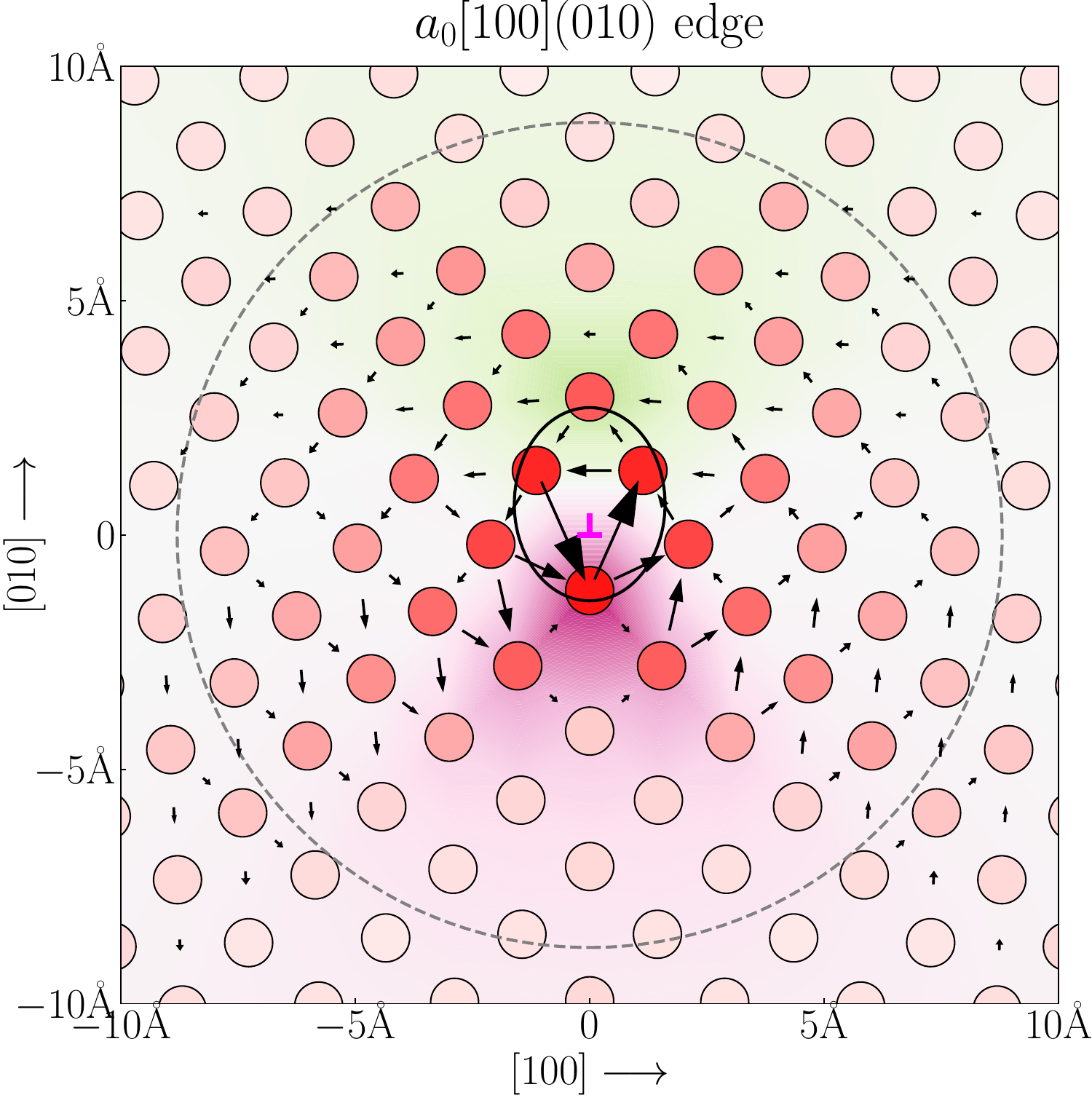}
        \includegraphics[width=\figwidth]{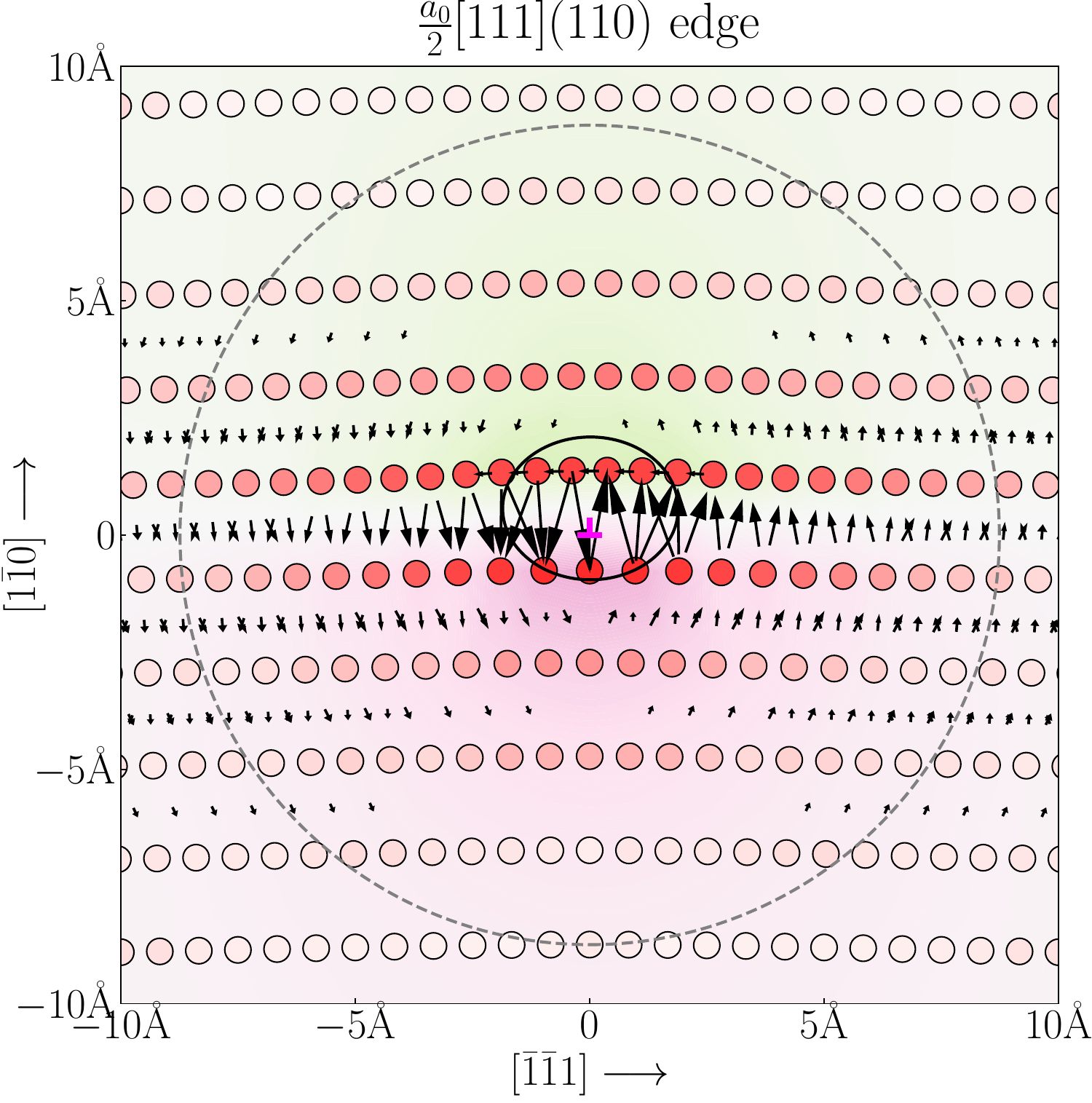}
    }
    \makebox[\textwidth][c]{%
        \includegraphics[width=\figwidth]{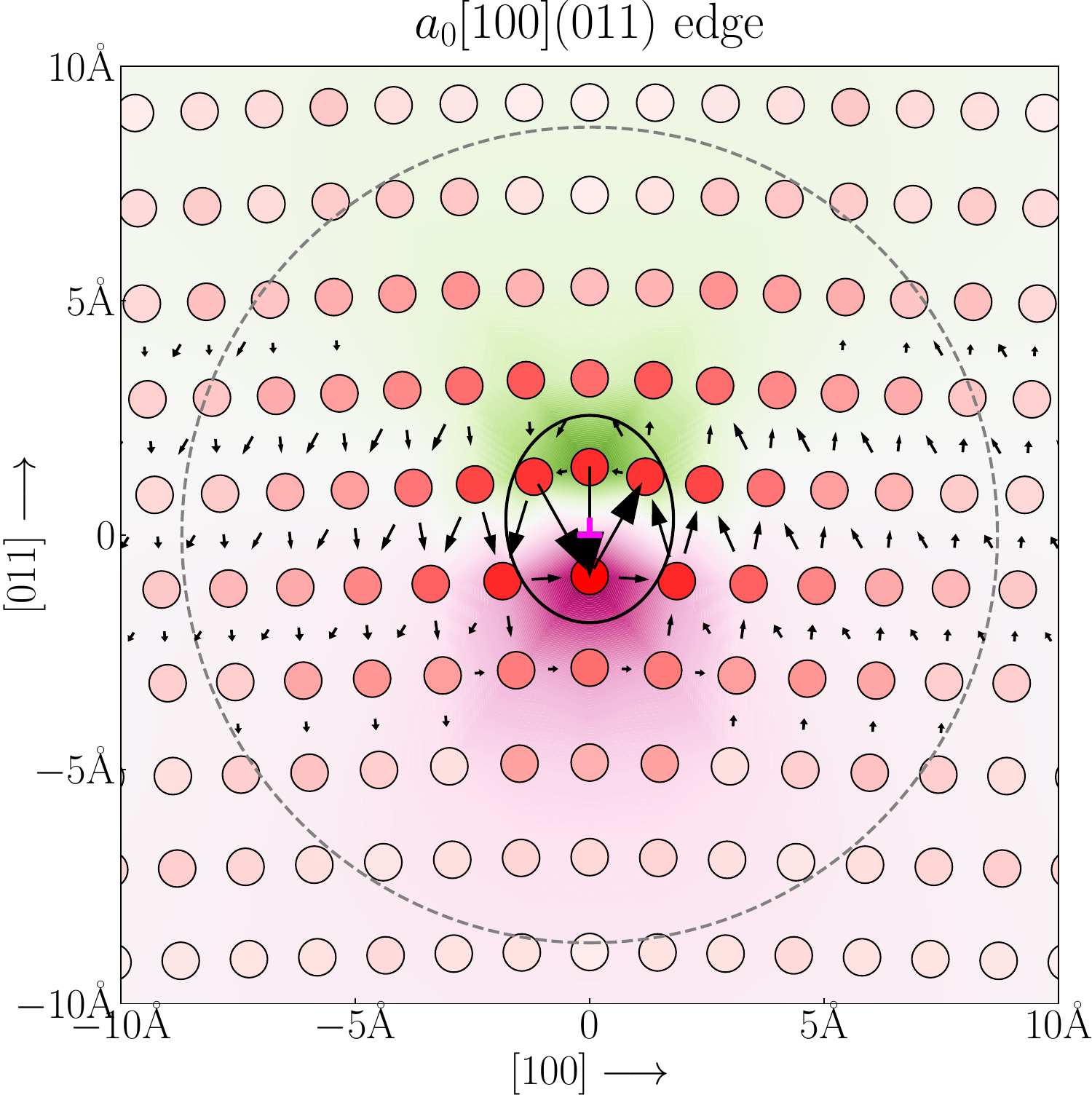}
        \includegraphics[width=\figwidth]{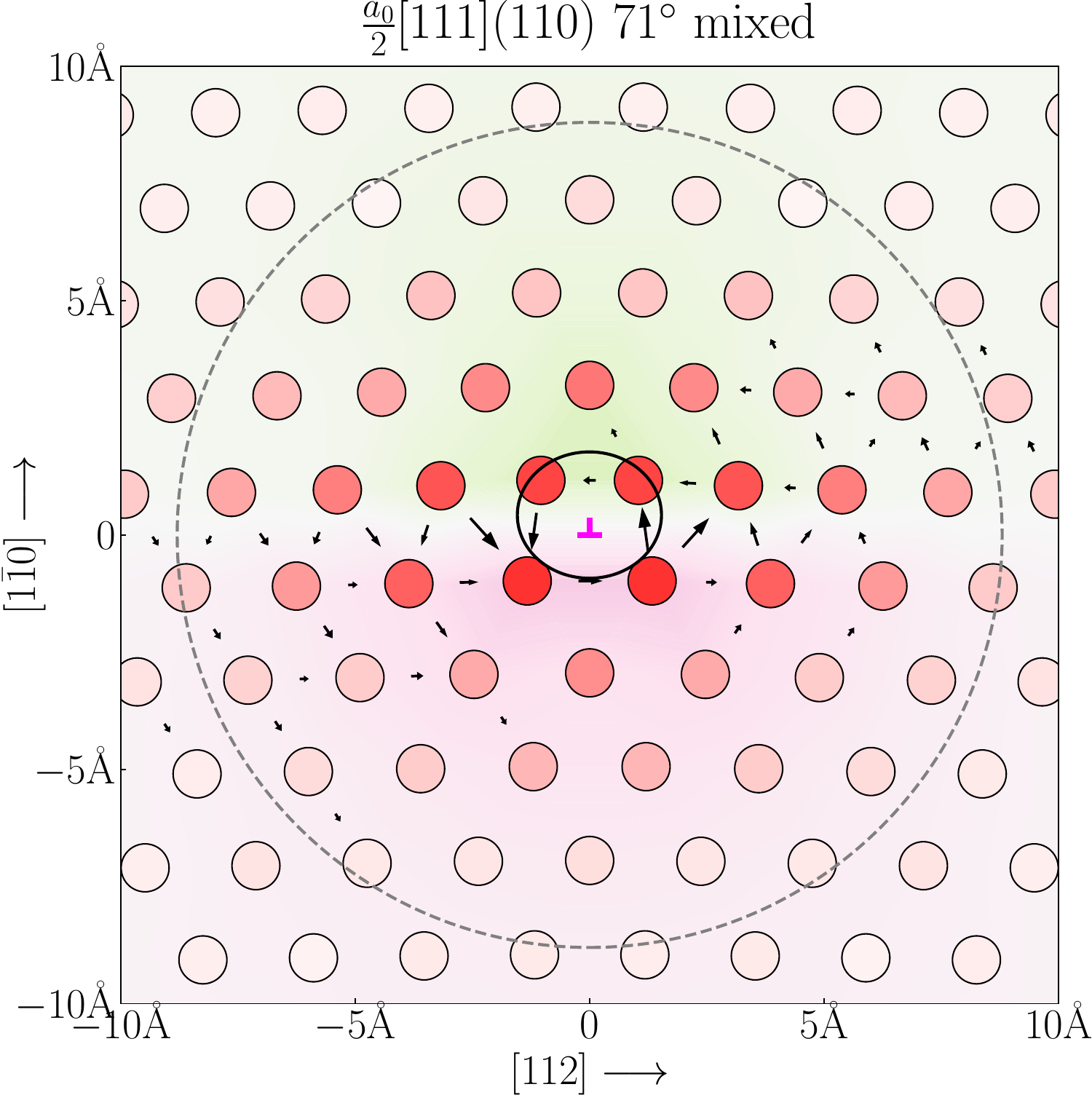}
    }
    \makebox[\textwidth][c]{%
        \qquad
        \includegraphics[width=\figwidth]{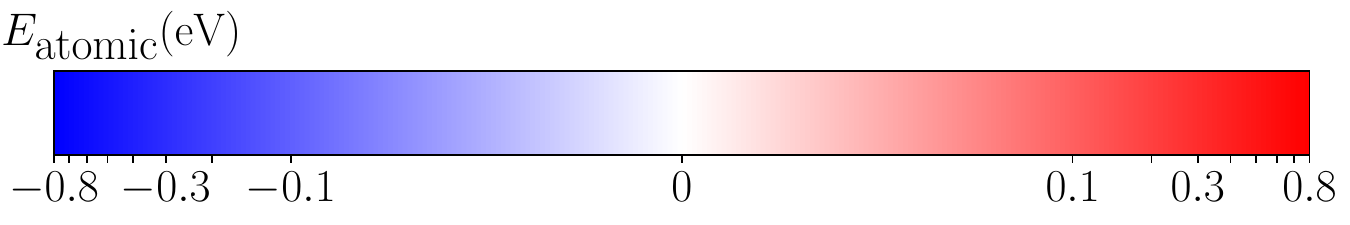}
        \includegraphics[width=\figwidth]{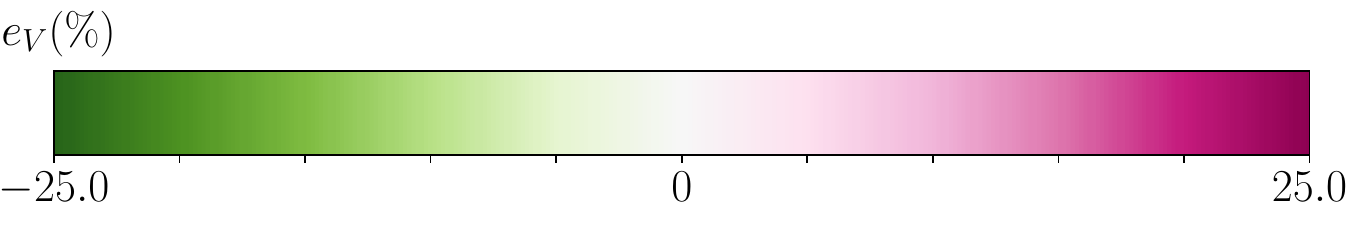}
    }
    \caption{DFT core structures of $a_0[100](010)$ edge, $\frac{a_0}{2}[\bar{1} \bar{1} 1](1\bar{1}0)$ edge, $a_0[100](011)$ edge and $\frac{a_0}{2}[111](1\bar{1}0)$ $71^\circ$ mixed dislocations in BCC Fe, prepared using DFT and LGF\cite{Fellinger2018}. The atoms are colored based on their EDM atomic energies, and the color contours show the distribution of volumetric strains. The gray dashed circle shows the boundary of the FBC region I. The black ellipses near the dislocation centers mark the widths of the core (listed in \Tab{corewidths}) from the center in and out of the slip planes, calculated using weighted deviation of atom coordinates, with weights defined by the magnitude of difference between EDM and anisotropic elastic atomic energies.}
    \label{fig:core-edm}
\end{figure*}

We compute the volumetric strains and differential displacements (DD) for visualization of the dislocation core properties. Each atomic site $s$ has $N_s$ nearest neighbor vectors $\{\mathbf{y}_1, \ldots, \mathbf{y}_{N_s}\}$ which correspond to bulk nearest neighbor vectors $\{\mathbf{x}_1,\ldots,\mathbf{x}_{N_s}\}$. The vectors in the dislocation geometry are related to the bulk vectors by the  local strain tensor $\boldsymbol{\epsilon}_s$ through $(\mathbf{1} + \boldsymbol{\epsilon}_s) \mathbf{x}_i = \mathbf{y}_i$, where $\mathbf{1}$ is the identity matrix and $i \in \{1, 2, \ldots, N_s \}$. The equality for all nearest neighbor vectors can only be satisfied in a least-squares sense, since $N_s$ equals 8 for most sites except those near the core center with large lattice distortions, leading to an overdetermined system with 24 equations for 9 unknown variables in $\boldsymbol{\epsilon}_s$. We associate the nearest neighbor vectors in the bulk $\mathbf{X}_s = [ {\mathbf{x}}_1^\intercal, \mathbf{x}_2^\intercal, \cdots, \mathbf{x}_{N_s}^\intercal]$ to the nearest neighbor vectors in the dislocation geometry $\mathbf{Y}_s = [\mathbf{y}_1^{\intercal}, \mathbf{y}_2^\intercal, \cdots, \mathbf{y}_{N_s}^\intercal]$ via linear transformation $(\mathbf{1} + \boldsymbol{\epsilon}_s) \mathbf{X}_s = \mathbf{Y}_s$, and solve it using least-squares method for the strain tensor
\begin{equation}
    \boldsymbol{\epsilon}_s = (\mathbf{X}_s \mathbf{X}_s^\intercal)^{-1} \mathbf{X}_s \mathbf{Y}_s^\intercal - \mathbf{1}.
\label{eqn:straintensor}
\end{equation}
The volumetric strain $\epsilon^V_s$ is then the trace, $\epsilon^V_s = \text{Tr}(\boldsymbol{\epsilon}_s)$. Because the volumetric strains $\epsilon^V_s$ are evaluated at discrete atomic sites in space, they are visualized in contour plots linearly interpolated between the atomic sites in \Fig{core-edm}. The core structures are visualized by the black arrows that represent the DD maps\cite{Vitek1970dd}. In a DD map, an arrow pointing between two atoms indicates the difference in their displacements along the direction of the Burgers vector, modulo $b/2$. The magnitude of the arrow is scaled so that a relative displacement of $b/2$ results in an arrow connecting the two atomic sites.

\begin{figure*}
    \makebox[\textwidth][c]{%
        \includegraphics[width=\figwidth]{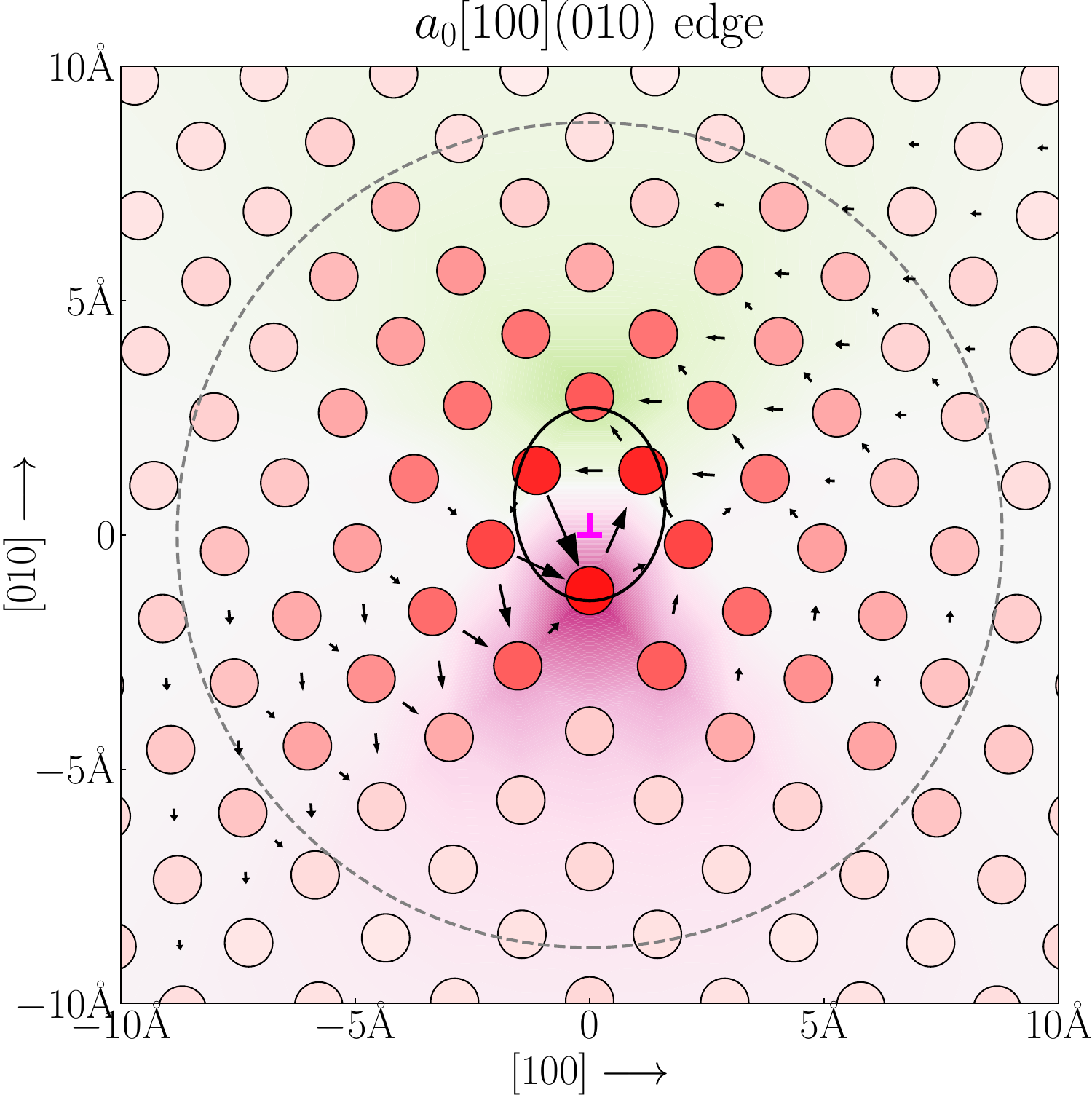}
        \includegraphics[width=\figwidth]{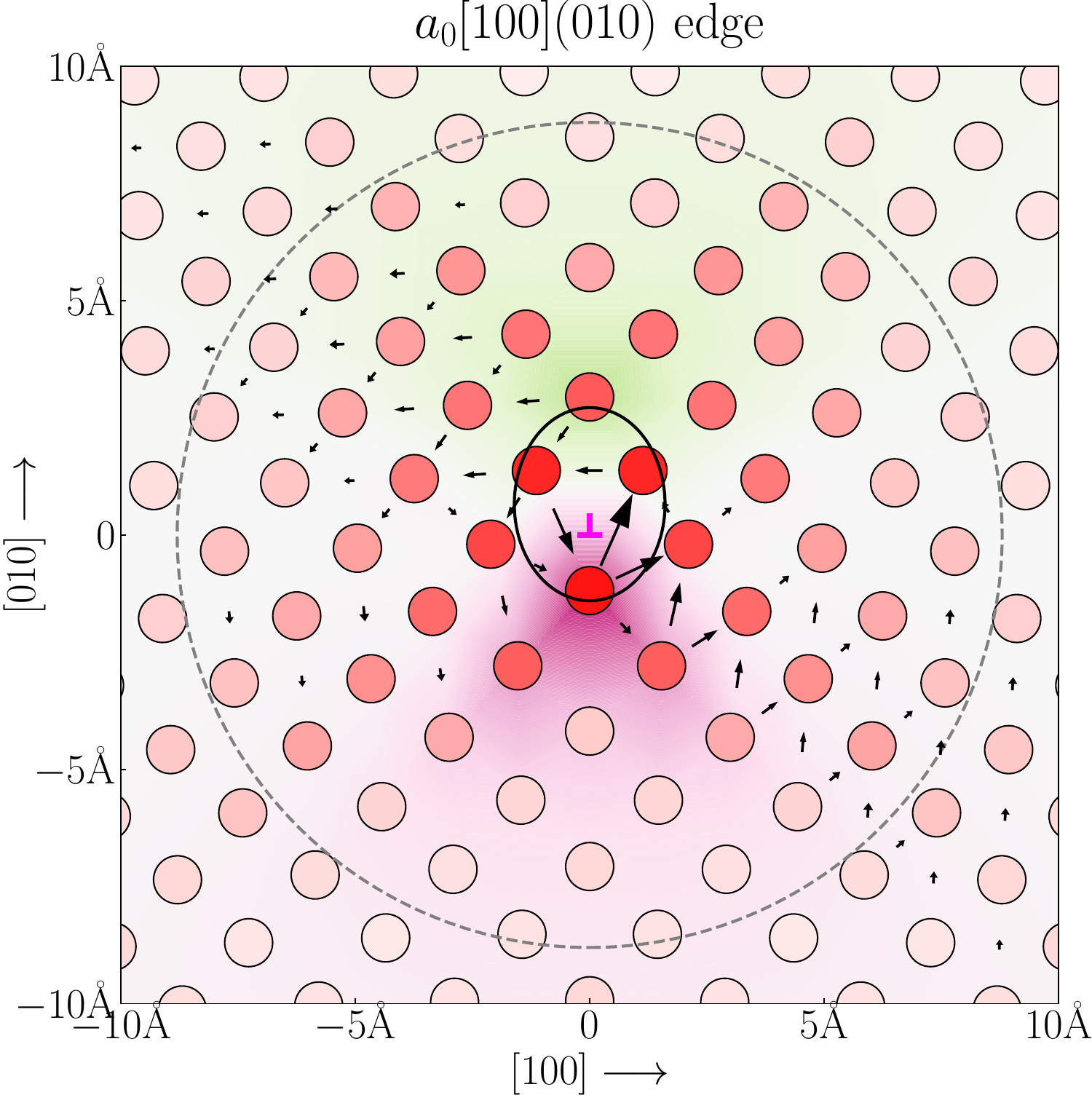}
    }
    \makebox[\textwidth][c]{%
        \qquad
        \includegraphics[width=\figwidth]{fig2e.pdf}
        \includegraphics[width=\figwidth]{fig2f.pdf}
    }
    \caption{Core spreading visualized by DD maps in the $(1\bar{1}0)$ (left) and $(110)$ (right) slip planes, showing that the $a_0[100](010)$ edge dislocation core spreads into two $\frac{a_0}{2}\langle 111 \rangle$ components.}
    \label{fig:core-edm-ddspread}
\end{figure*}

The region I atoms in the dislocation core have atomic energies that match the distribution of the volumetric strains and the core structures revealed by the DD map, as is shown in \Fig{core-edm}. For each dislocation, the region above (below) the slip plane, which is located at $n=0$ and perpendicular to the $n$-axis, is compressive (tensile), as shown by the volumetric strains. Large volumetric strains coincide with high EDM atomic energies near the core centers, indicating that local tension and compression both increase the local atomic energies. The DD maps show compact cores for all the dislocations with
DDs large near the core center and smaller when moving away from it. Near the slip planes, the DDs are greater compared to those of the atoms above and below, reflecting the lattice distortions caused by the slips. This also coincides with the high EDM atomic energies near the slip planes in all the dislocations. For the $a_0[100](010)$ edge dislocation, the core shows spreading in the $(1\bar10)$ and $(110)$ slip planes, with dislocation content remaining from the reaction in \Eqn{reaction}, which corresponds to the relatively large DDs along the diagonal $\pm 45^\circ$ directions from the $-n$ axis. EDM shows distribution of higher atomic energies in the same directions, capturing the effect of the core spreading. This can be seen more clearly in \Fig{core-edm-ddspread}, which separately shows the DD maps for $\frac{a_0}{2}[111]$ and $\frac{a_0}{2}[1\bar1\bar1]$ Burgers vectors. Despite the favorability of the reaction, the dislocation core structure shows the remnants of two dislocations involved despite being created with a perfect $a_0[100]$ Burgers vector.

\begin{figure*}[htbp]
    \makebox[\textwidth][c]{%
        \includegraphics[width=\halfwholefigwidth]{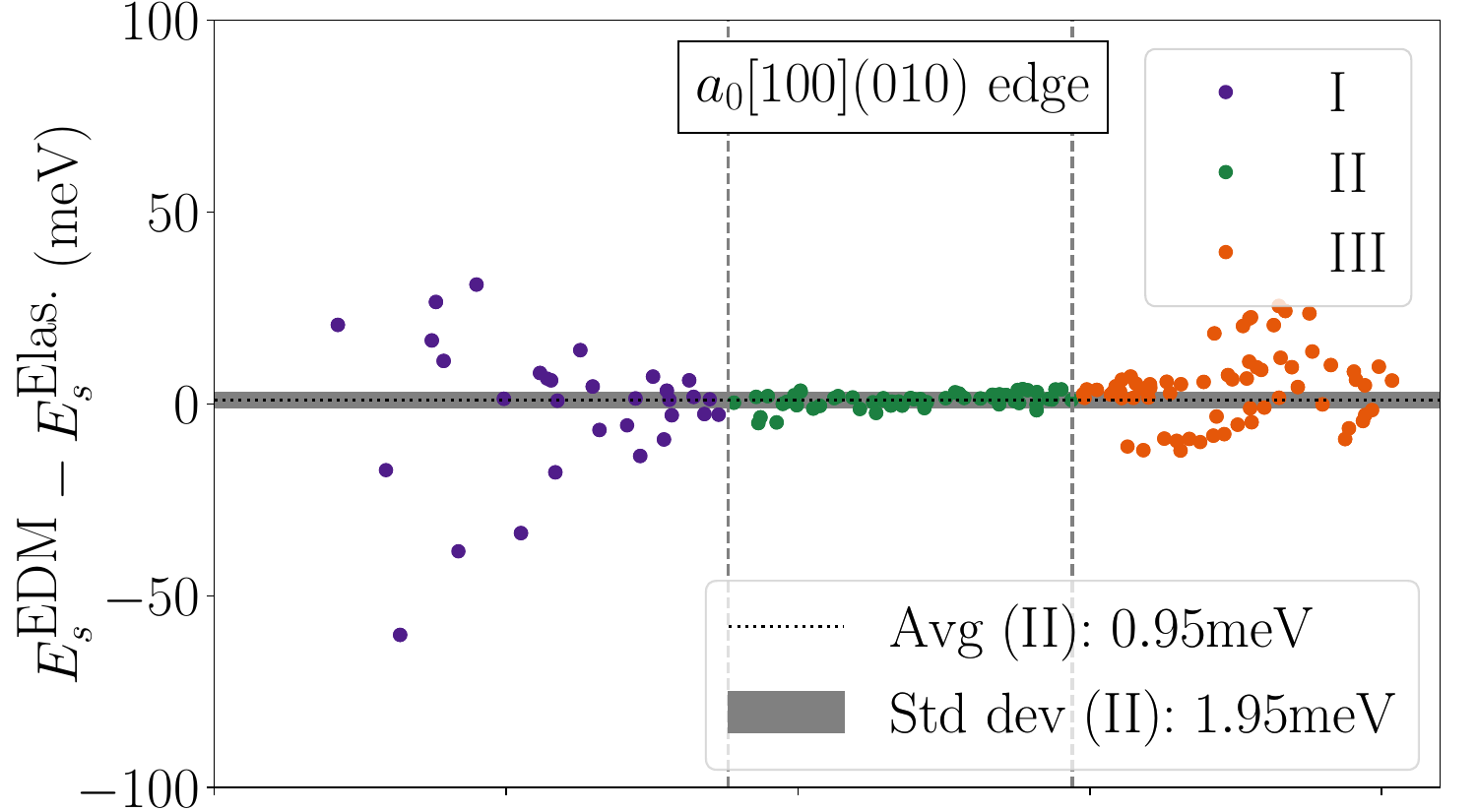}
        \includegraphics[width=\halfwholefigwidth]{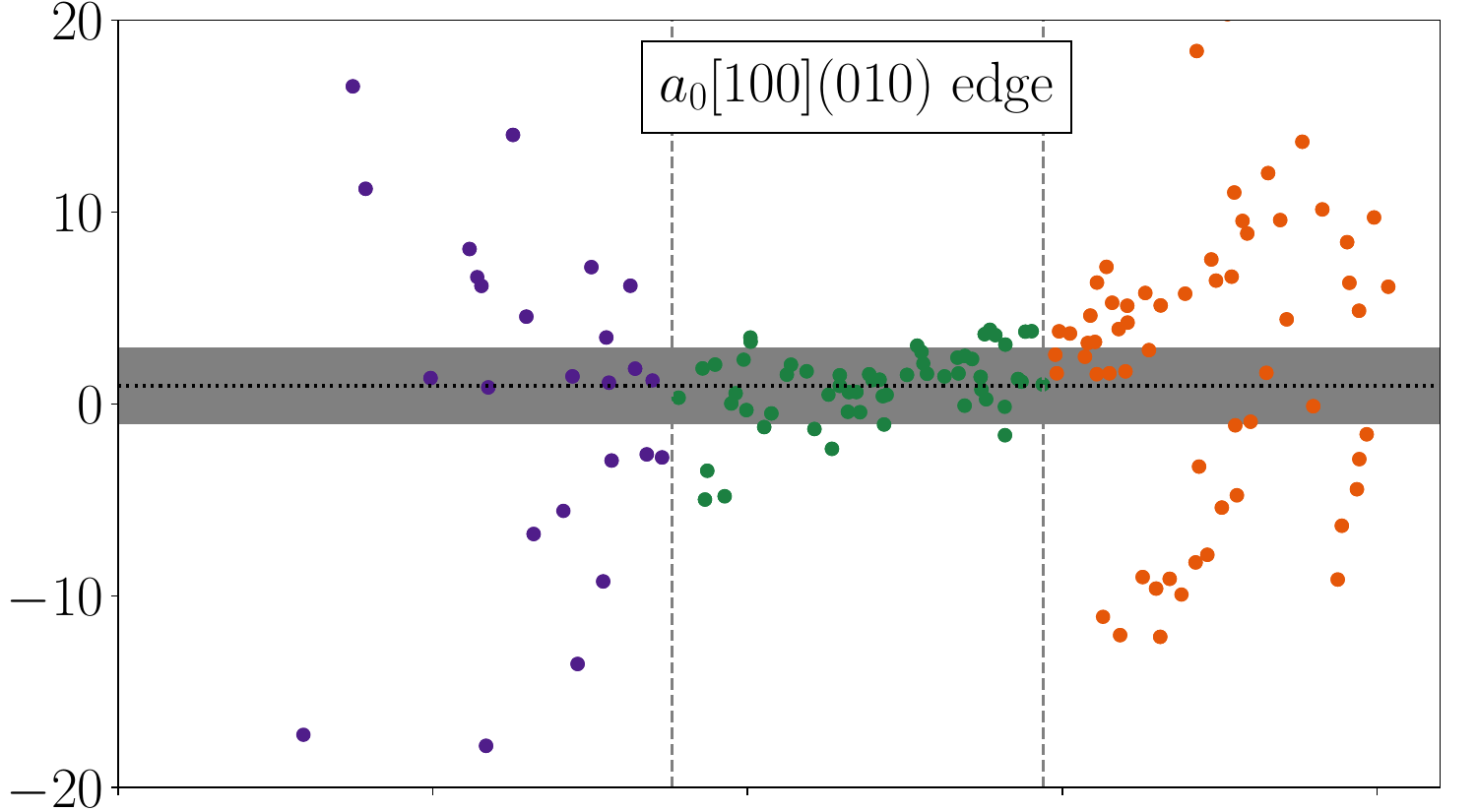}
    }
    \makebox[\textwidth][c]{%
        \includegraphics[width=\halfwholefigwidth]{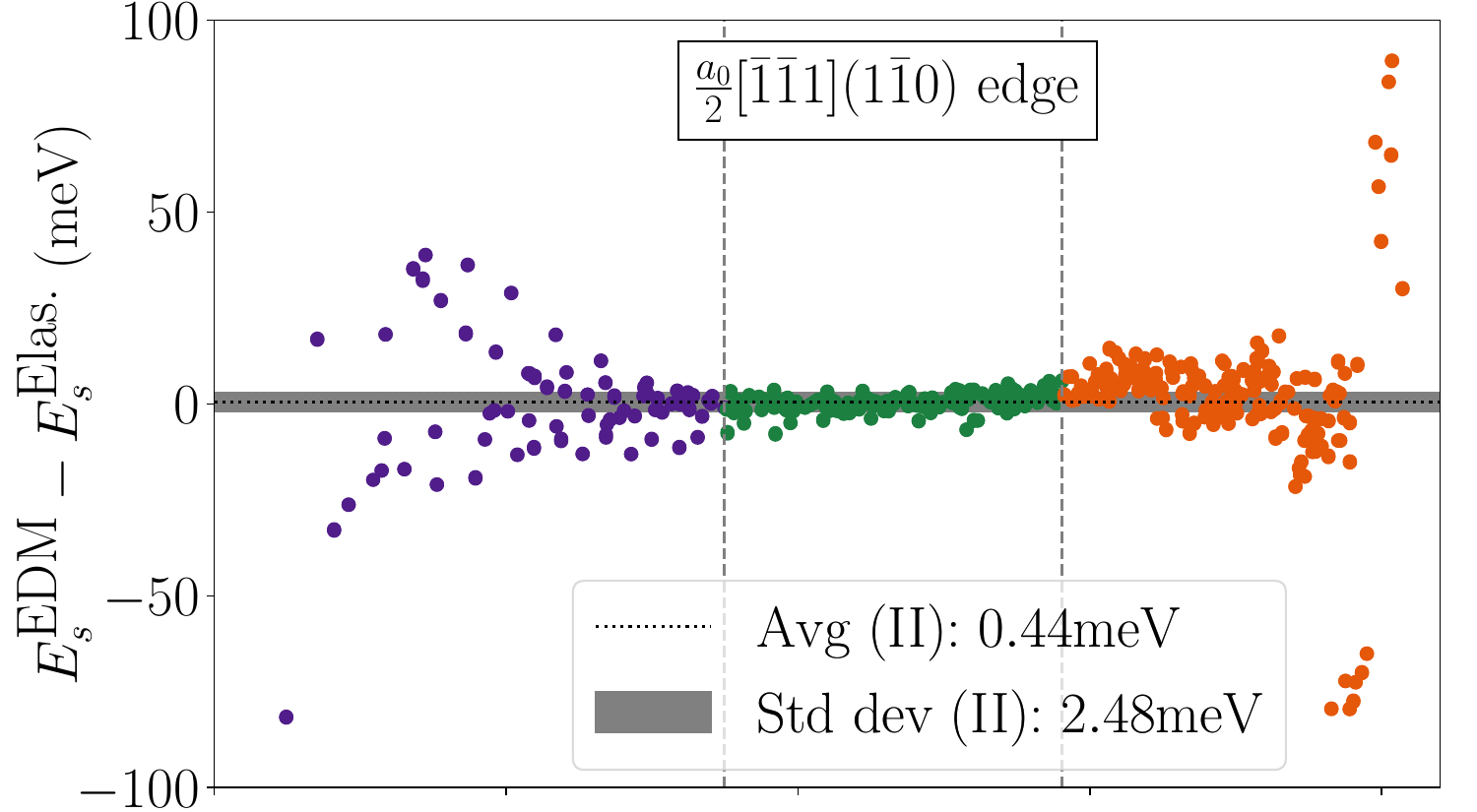}
        \includegraphics[width=\halfwholefigwidth]{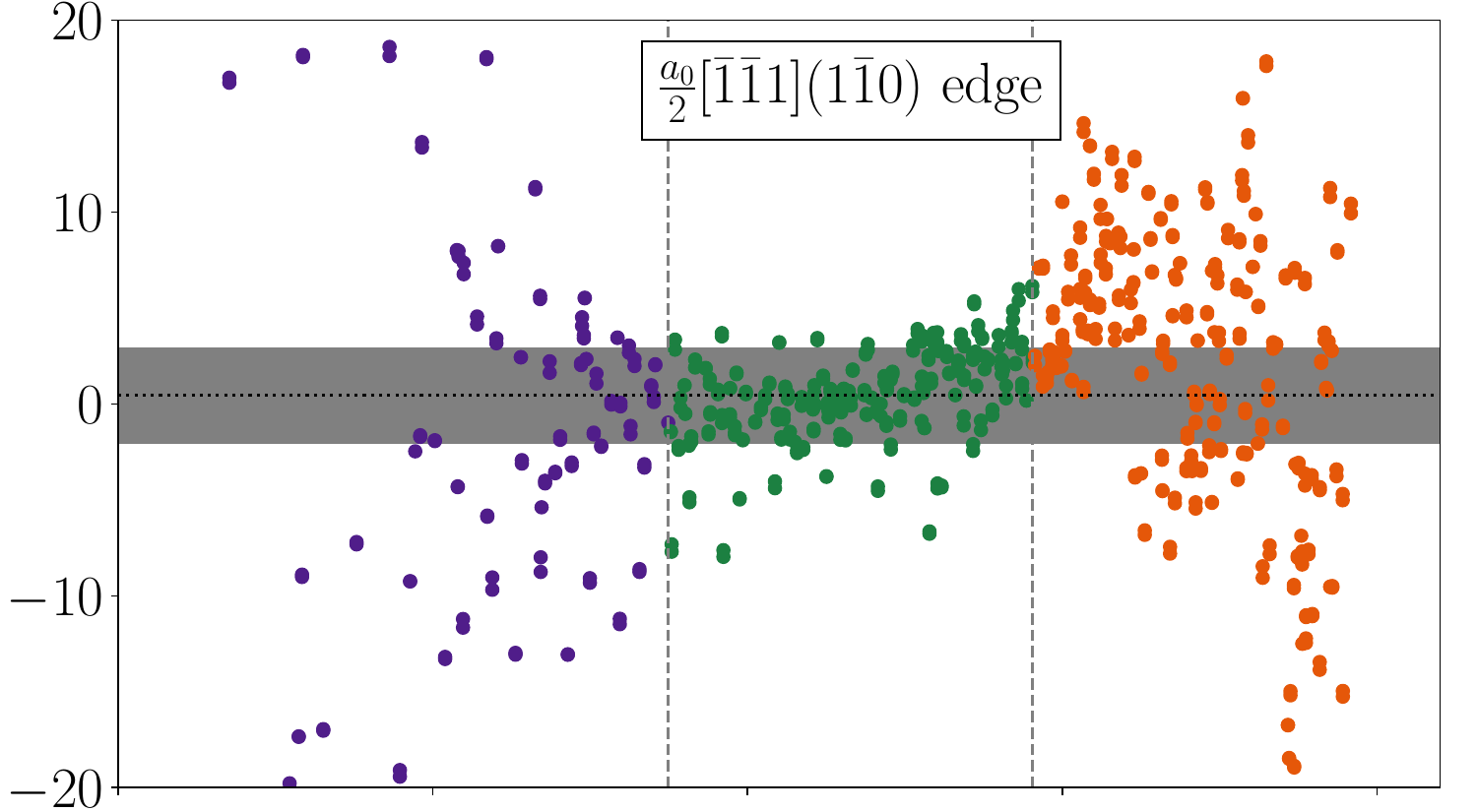}
    }
    \makebox[\textwidth][c]{%
        \includegraphics[width=\halfwholefigwidth]{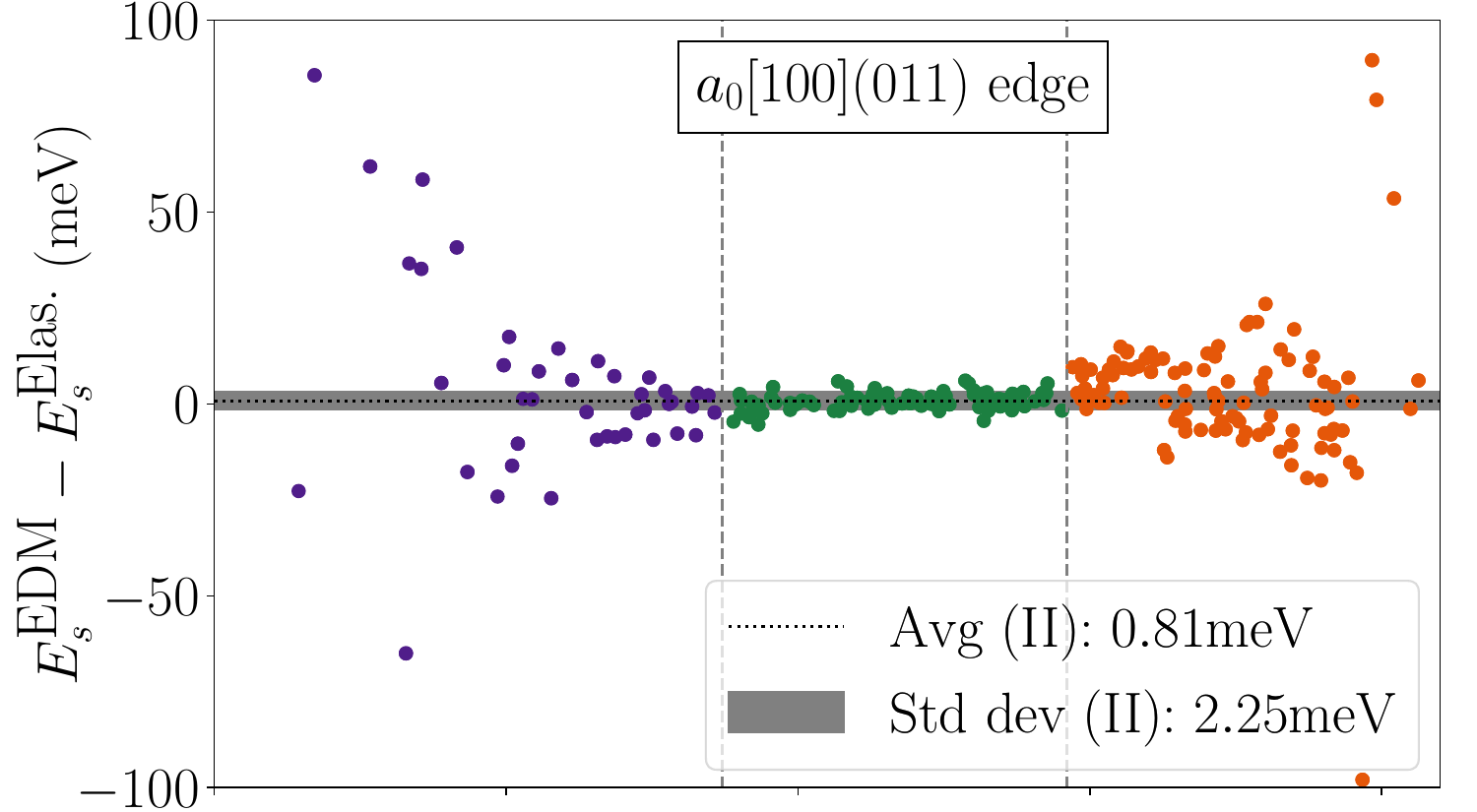}
        \includegraphics[width=\halfwholefigwidth]{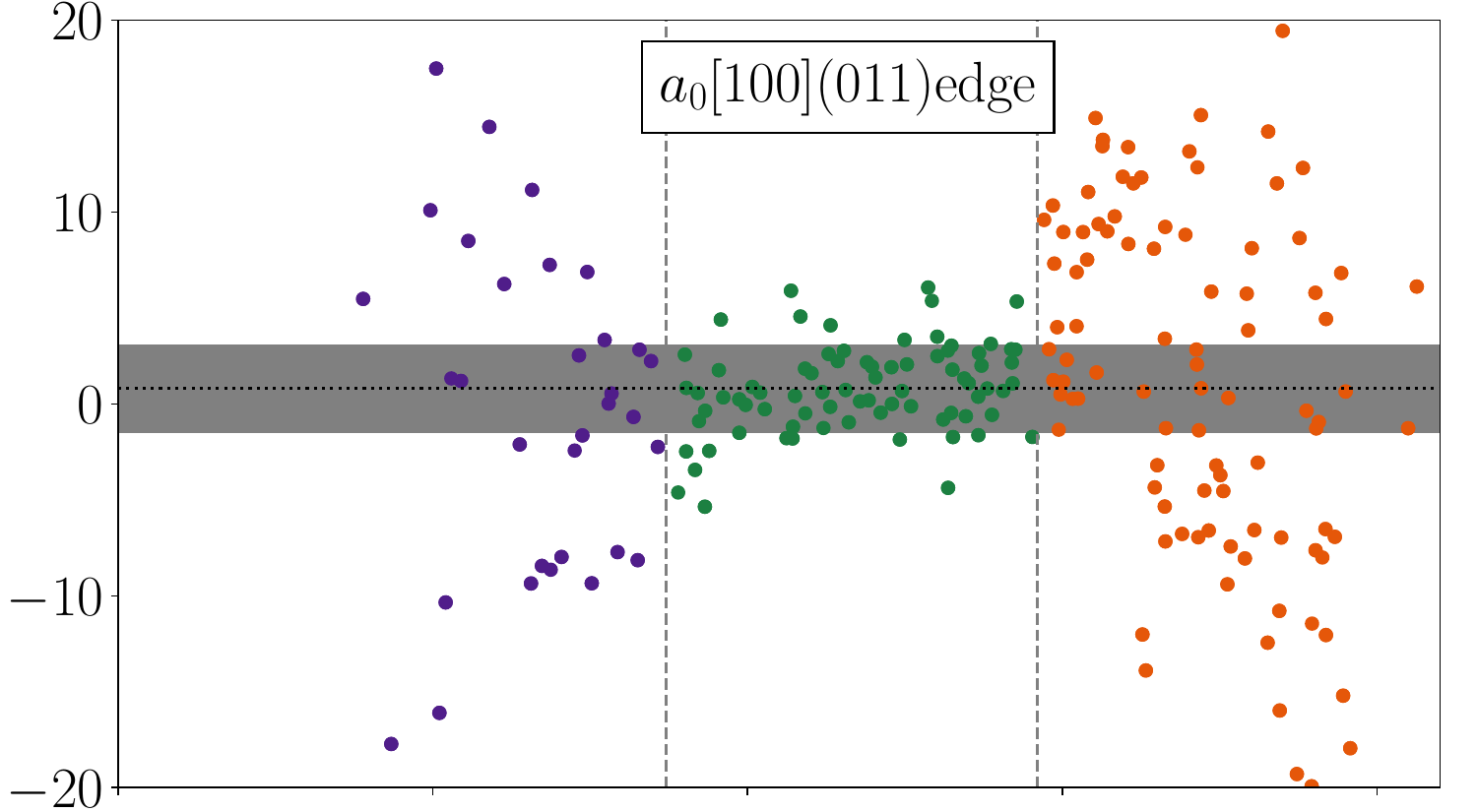}
    }
    \makebox[\textwidth][c]{%
        \includegraphics[width=\halfwholefigwidth]{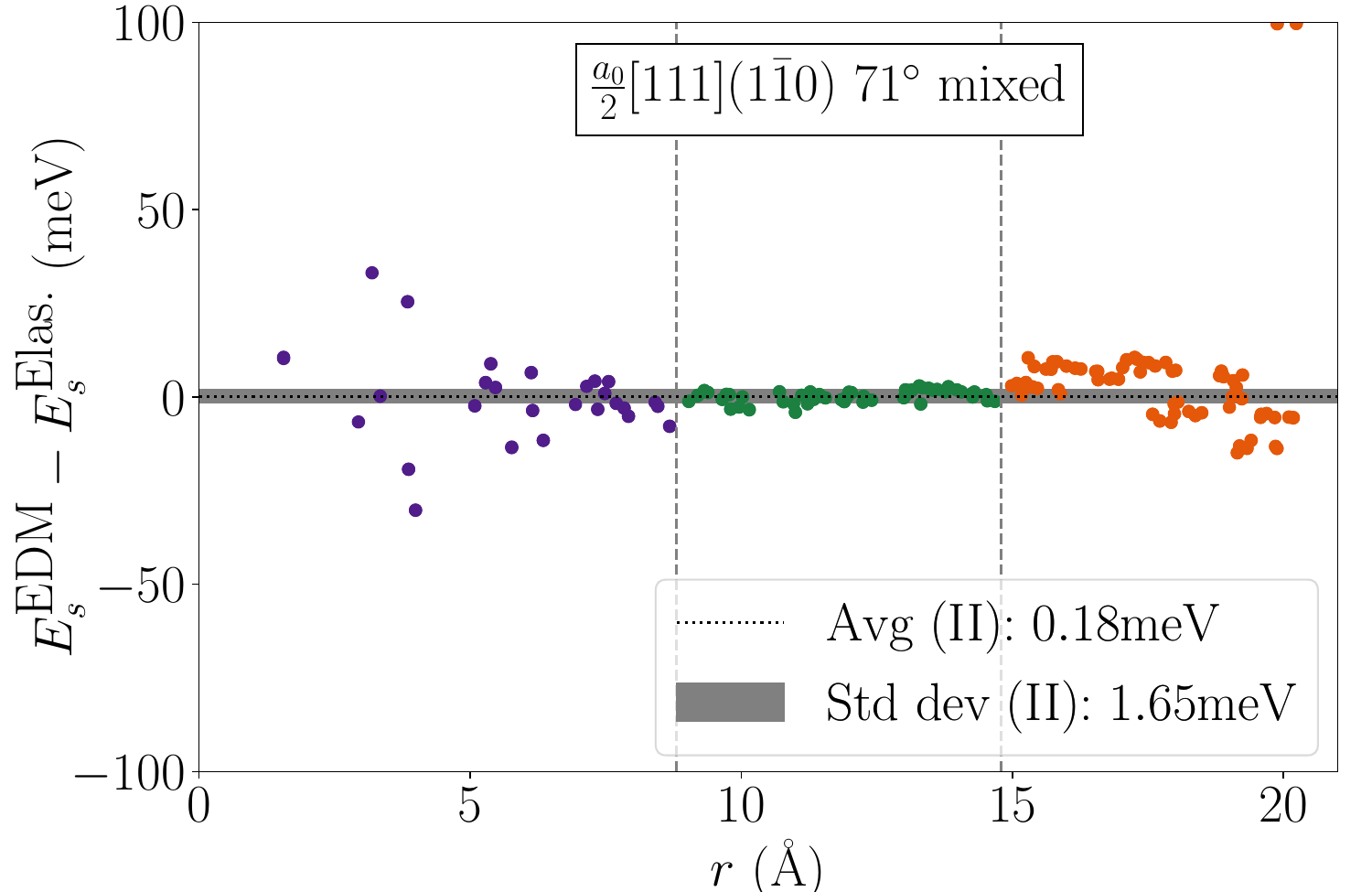}
        \includegraphics[width=\halfwholefigwidth]{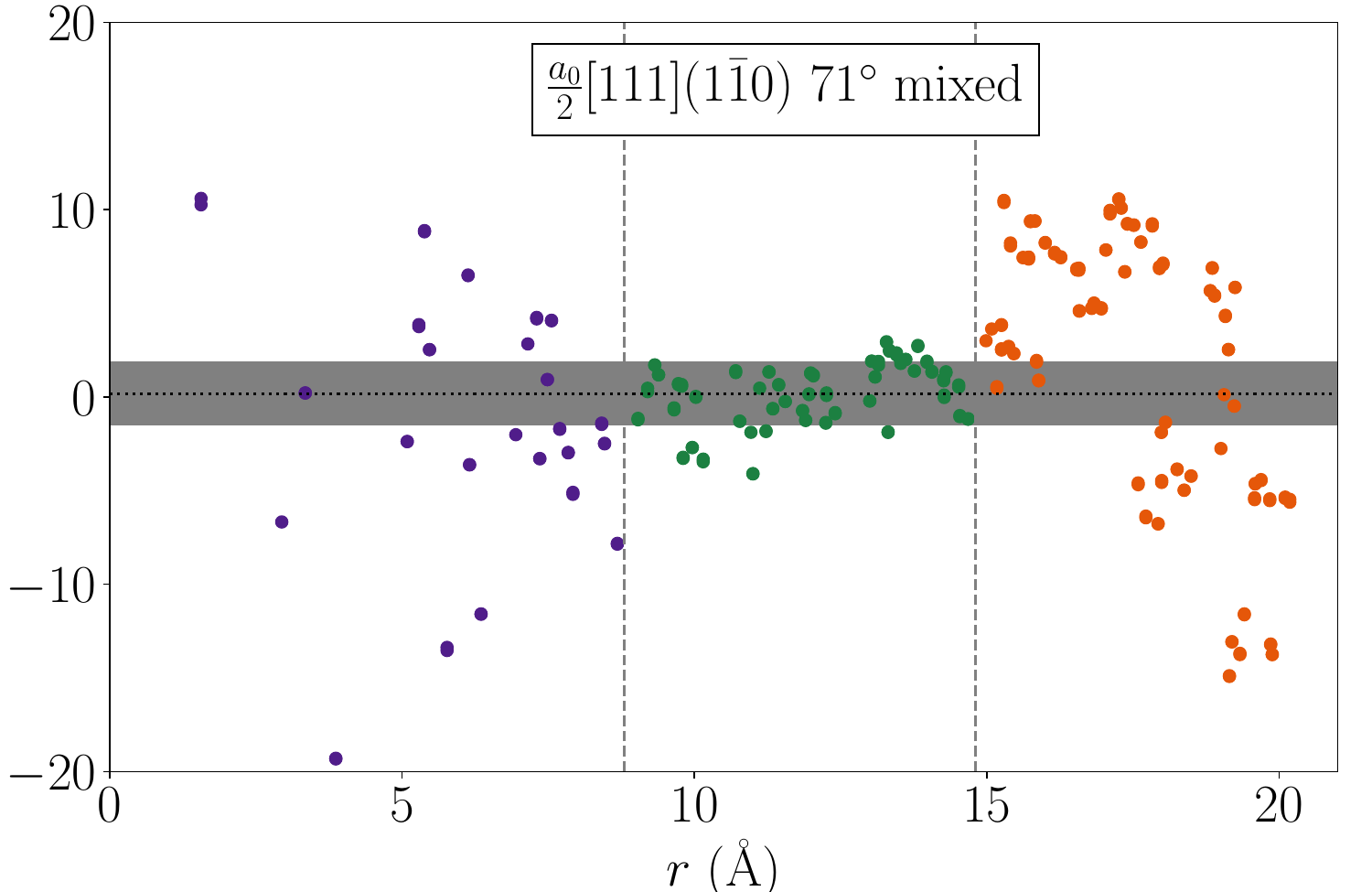}
    }
    \caption{Difference between EDM atomic energies and anisotropic elastic energies (left column) and with a smaller energy range (right column) that highlight region II and plotted as function of $R$, the distance to the dislocation center for elasticity. Mean and standard deviations of the energy difference for region II atoms are marked by the horizontal lines and shaded grey areas, and their numbers are shown in the figures in the right column.}
    \label{fig:DeltaE}
\end{figure*}

The EDM energies agree well with the predictions of anisotropic elasticity in region II between the core and the free surface where atomic displacements are small, while capturing the nonlinearity behavior in region I near the core, as is illustrated in \Fig{DeltaE}. The anisotropic elastic energies are calculated from the 4-rank elasticity tensor $\mathbf{C}$ and the 2-rank strain tensor $\boldsymbol{\epsilon}_s$ with two double dot products
\begin{equation}
    E^\text{elas}_s = \frac{V_0}{2} \boldsymbol{\epsilon}_s : \mathbf{C} : \boldsymbol{\epsilon}_s
    \label{eqn:Eelas}
\end{equation}
where $V_0$ is the volume of the primitive cell. The elasticity tensor $\mathbf{C}$ can be determined from three elastic constants based on the symmetry of BCC crystal, $C_{11}$, $C_{22}$ and $C_{44}$, which take the values of 277.5 GPa, 147.7 GPa and 98.1 GPa, respectively\cite{Fellinger2018} in this work. We calculate the difference between EDM and anisotropic elastic energies for each atom $s$
\begin{equation}
    \Delta E_s = E^\text{EDM}_s - E^\text{Elas.}_s
    \label{eqn:Ediff},
\end{equation}
which turns out to be small in region II, with average value less than $1$ meV/atom and standard deviation at $2.5$ meV/atom or less, as shown in \Fig{DeltaE}. This indicates that EDM well matches the energy prediction of anisotropic elasticity at small strains. The energy differences diverge in region I due to the failure of anisotropic elasticity in the highly distorted cores. Region III also shows large energy differences but due to the spurious energy contributions from the free surface, and are excluded from further analyses. In region II, the innermost atoms close to region I and the outermost atoms close to region III are well-behaved with the energy differences being small and similarly distributed compared to those of the atoms close to the center of region II. This is an illustration that the geometries are sufficiently large for LGF and the boundaries for the regions are reasonably chosen.

\begin{figure*}[htbp]{%
    \includegraphics[width=\figwidth]{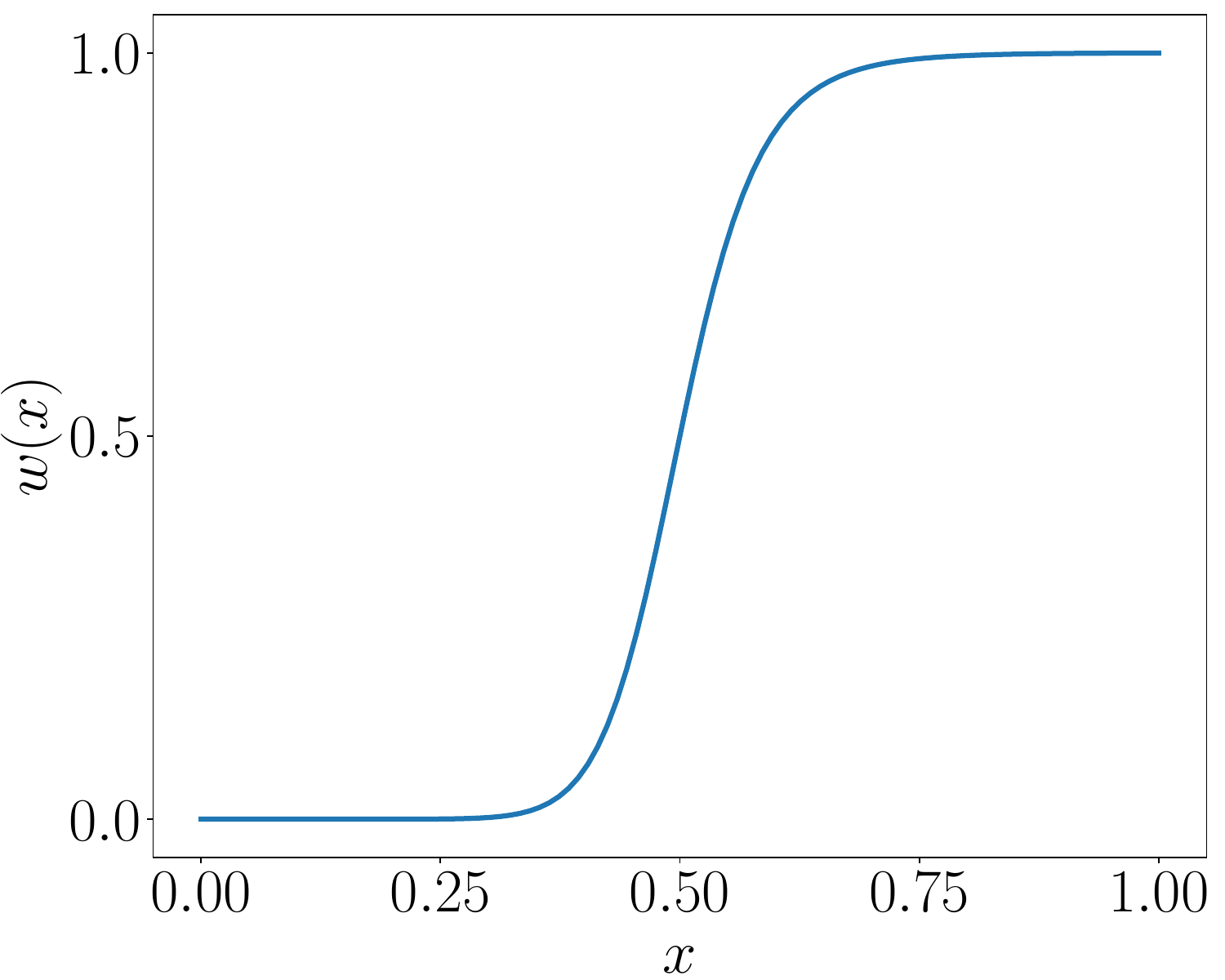}
    \caption{The sigmoid weight function $w(x)$ defined by \Eqn{weight} used for the calculations of core widths, which smoothly increases from 0 at $x=0$, reaching 0.5 at $x=\eta=0.5$, and approaches 1 for large $x$.}
    \label{fig:sigmoid_weight}
}
\end{figure*}

In the core of a dislocation, linear elasticity theory fails to predict atomic displacements; we therefore estimate the size of the cores from the range of large deviations of EDM energies from the elastic energies. We consider an atom to be in the core if $|\Delta E_s| > \frac12 E_s^\text{elas}$. In lieu of a hard cutoff for selecting the core atoms, we use a sigmoid function that smoothly transitions from 0 to 1,
\begin{equation}
    w(x) =  \frac{1}{1 + \exp{ - k \ln \left( \frac{x}{\eta} \right) } } = \frac{x^k}{x^k + \eta^k} \\
    \label{eqn:weight}
\end{equation}
with $\eta = 1/2$ and $k=12$, as shown in \Fig{sigmoid_weight}. We define the core widths as weighted standard deviations of the site coordinates, with the weight dependent on the relative energy difference $| \Delta E_s / E_s^\text{elas} |$ for atoms in regions I and II. Therefore, the widths in ($\delta m$) and perpendicular to ($\delta n$) the dislocation slip plane can be expressed as
\begin{equation}
    \begin{aligned}
        \delta m &:= \sqrt{ \sum_{s} (m_s - \bar{m})^2 w \left( \left| \frac{\Delta E_s}{E^\text{elas}_s} \right| \right) } \\
        \delta n &:= \sqrt{ \sum_{s} (n_s - \bar{n})^2 w \left( \left| \frac{\Delta E_s}{E_s^\text{elas}} \right| \right) } \\
\end{aligned},
    \label{eqn:corewidths}
\end{equation}
where $\bar{m}$ and $\bar{n}$ are weighed average positions
\begin{equation}
    \begin{aligned}
        \bar{m} &= \sum_{s} m_s w \left( \left| \frac{\Delta E_s}{E_s^\text{elas}} \right| \right) \\
        \bar{n} &= \sum_{s} n_s w \left( \left| \frac{\Delta E_s}{E_s^\text{elas}} \right| \right)
    \end{aligned}.
\label{eqn:avgpos}
\end{equation}
In other words, the $(\bar{m}, \bar{n})$ and $(\delta m, \delta n)$ are the first and second moments of the energy-dependent weight function $w\left( \left| \frac{\Delta E_s}{E_s^\text{elas}} \right| \right)$, respectively. The weights are normalized so that they sum to 1 over all atoms.

\Tab{corewidths} collects the calculated core widths, which the dislocation figures present as black ellipses. The ellipses are made with their axes defined by $\delta m$ and $\delta n$ in \Eqn{corewidths} and centered at $(\bar{m}, \bar{n})$ determined from \Eqn{avgpos}. In general, the ellipses are centered near the dislocation centers and cover the areas with the maximum DDs, indicating that the core widths are able to illustrate the core structures from the energy perspective. In all the cases, the core widths are less than 4 \AA, implying compact cores for all the dislocations. The core widths are also computed as the first and second moments of the Nye tensor distributions by Fellinger \textit{et al.}\cite{Fellinger2018}, which are reported to range from 3.0 \AA\ to 4.7 \AA\ in the $m$-direction and $n$-direction, similar to the magnitude of the widths from our calculations. The Nye tensor component $\alpha_{31}$, which corresponds to the local Burgers vector density in the $m$-direction and is the major non-zero edge component in all the dislocations, gives $m$-to-$n$ aspect ratios of 0.96 and 1.44 for the core widths in the $a_0[100](010)$ edge and $\frac{a_0}{2}[\bar{1}\bar{1}1](1\bar{1}0)$ edge dislocations, respectively. These values qualitatively agree with the corresponding aspect ratios of 0.81 and 1.27 for the core widths calculated from site energies in this work. The $\frac{a_0}{2}[111](1\bar{1}0)$ $71^\circ$ mixed dislocation has both edge and screw Nye tensor components, $\alpha_{31}$ and $\alpha_{33}$, leading to similar aspect ratios of 1.21 and 1.34, while the core widths calculated from energy reflects a combined influence of both edge and screw components and leads to 1.18 in the aspect ratio. The $\delta m$ widths, however, are systematically smaller when calculated using energy difference data than when calculated using Nye tensor components, especially for the $a_0[100](011)$ edge dislocation, where the Nye tensor component $\alpha_{31}$ gives an $m$ width of 4.31 \AA\ while our energy difference data yield 3.26 \AA. This shows that the Nye tensor components have larger dispersion compared to our criterion for core atoms, along the $m$-axis. 

\begin{table*}[htbp]
    \caption{Core widths $\delta m \times \delta n$ calculated using different geometries and methods, shown in \AA. Columns DFT (EDM), EAM and GAP list the core widths calculated from \Eqn{corewidths} with energy data in this work, for geometries made using DFT, the EAM potential and the GAP potential, respectively. Column DFT (Nye tensor) lists the core widths calculated as the second moments of the Nye tensor components\cite{Fellinger2018}, for the same DFT geometries. The Nye tensor components $\alpha_{31}$, $\alpha_{32}$ and $\alpha_{33}$ correspond to the local Burgers vector density in the $m$, $n$ and $t$ directions, respectively.}
    \begin{tabular}{c@{\quad}c@{\quad}c@{\quad}c@{\quad}c@{\quad}}
        \hline \hline
        & DFT (EDM)  & DFT (Nye tensor)\cite{Fellinger2018} & EAM & GAP \\
        \hline
        \multirow{2}{*}{$a_0[100](010)$ edge} & \multirow{2}{*}{$3.12 \times 3.85$} & $3.78 \times 3.92$, $\alpha_{31}$ & \multirow{2}{*}{$5.19 \times 3.65$} & \multirow{2}{*}{$4.12 \times 3.54$} \\
        & & $4.69 \times 3.04$, $\alpha_{32}$ & & \\
        $\frac{a_0}{2}[\bar{1}\bar{1}1](1\bar{1}0)$ edge & $3.85 \times 3.02$ & $4.33 \times 3.00$, $\alpha_{31}$ & $5.28 \times 1.17$ & $3.89 \times 3.38$ \\
        $a_0[100](011)$ edge & $3.26 \times 3.86$ & $4.31 \times 3.38$, $\alpha_{31}$ & $3.47 \times 3.94$ & $3.62 \times 4.14$ \\

        \multirow{2}{*}{$\frac{a_0}{2}[111](1\bar{1}0)$ $71^\circ$ mixed} & \multirow{2}{*}{$3.10 \times 2.63$} & $3.97 \times 3.28$, $\alpha_{31}$ & \multirow{2}{*}{$2.85 \times 1.54$} & \multirow{2}{*}{$3.01 \times 3.46$} \\
        & & $4.41 \times 3.30$, $\alpha_{33}$ & & \\
        \hline \hline
    \end{tabular}
    \label{tab:corewidths}
\end{table*}

\begin{figure*}[htbp]
    \makebox[\textwidth][c]{%
        \includegraphics[width=\figwidth]{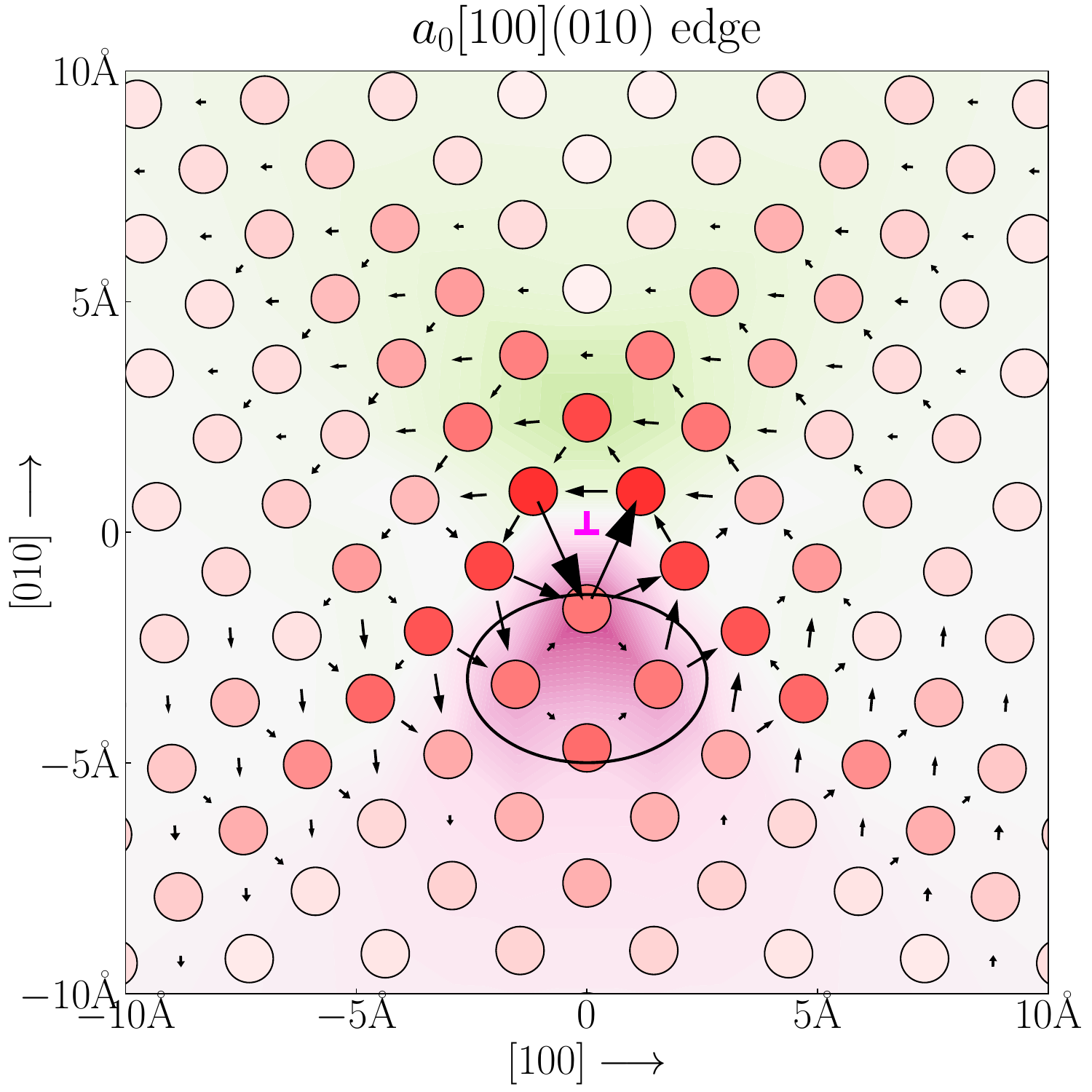}
        \includegraphics[width=\figwidth]{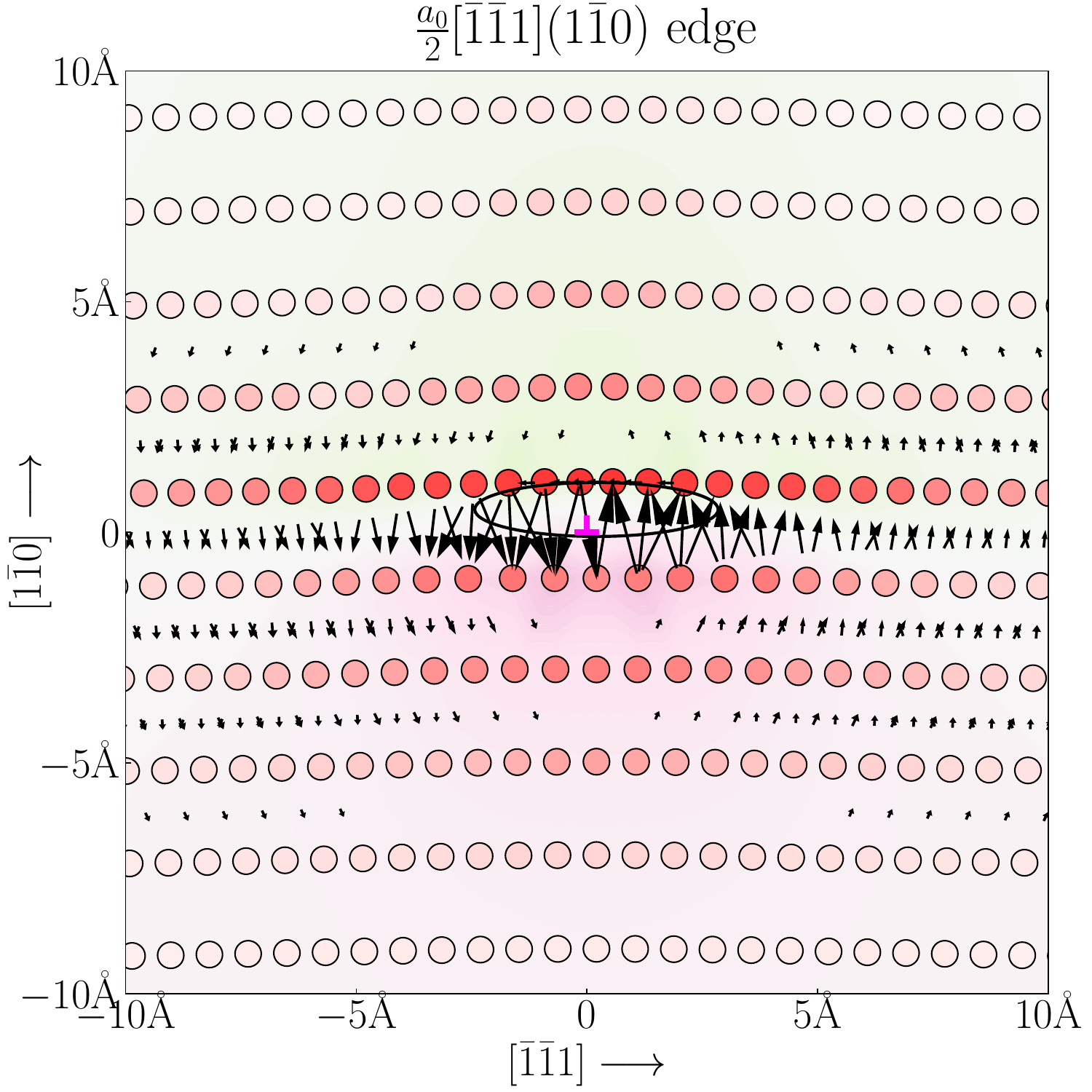}
    }
    \makebox[\textwidth][c]{%
        \includegraphics[width=\figwidth]{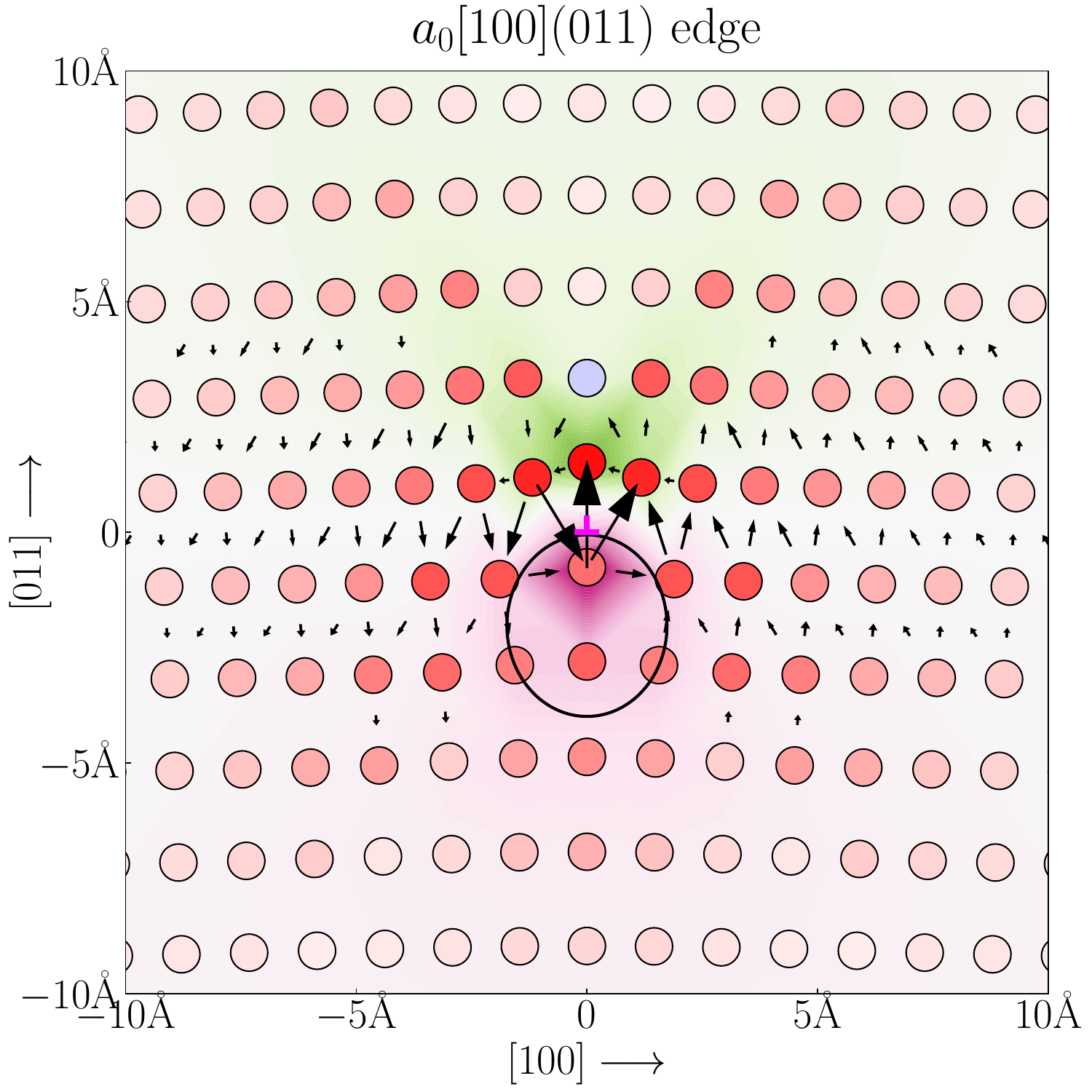}
        \includegraphics[width=\figwidth]{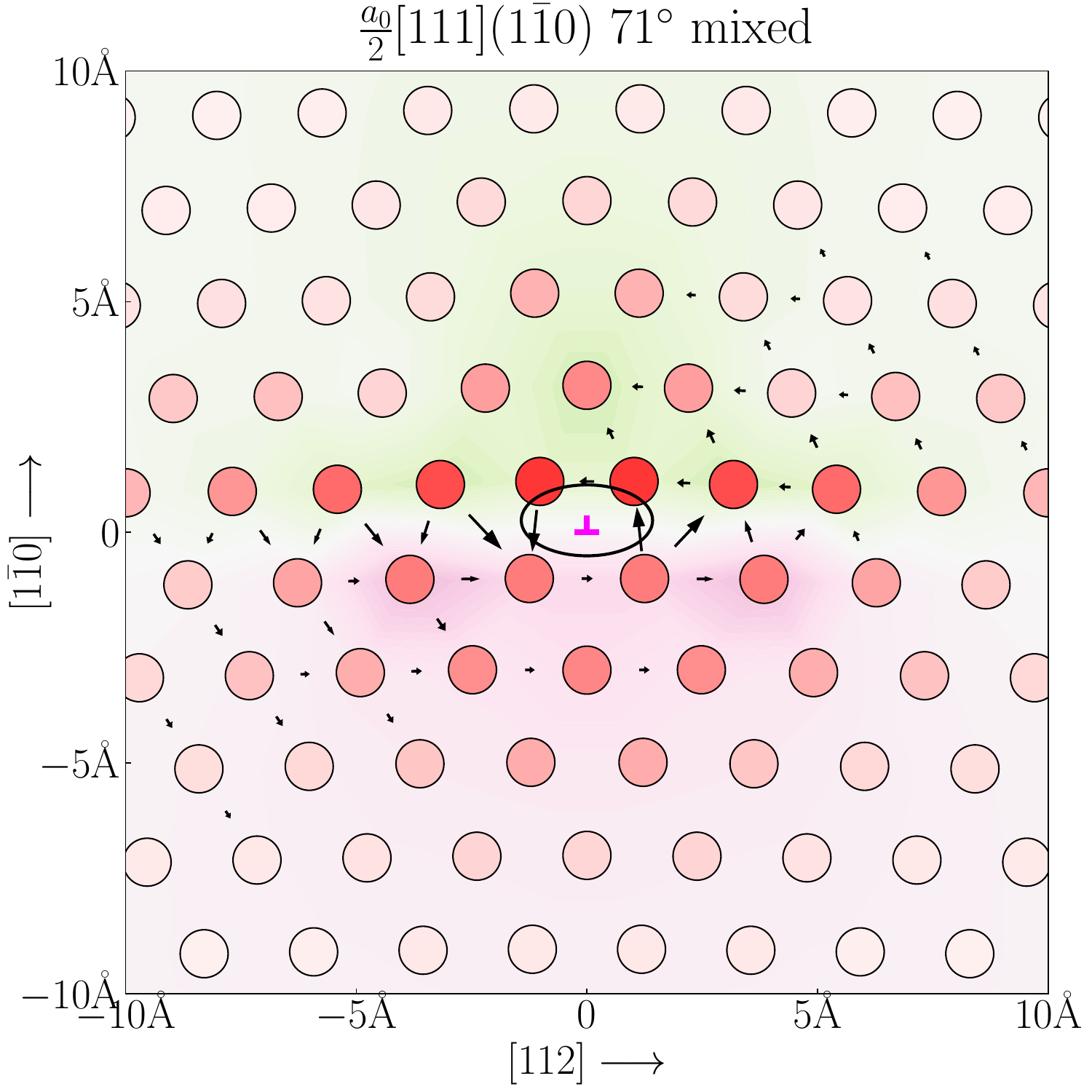}
    }
    \makebox[\textwidth][c]{%
        \qquad
        \includegraphics[width=\figwidth]{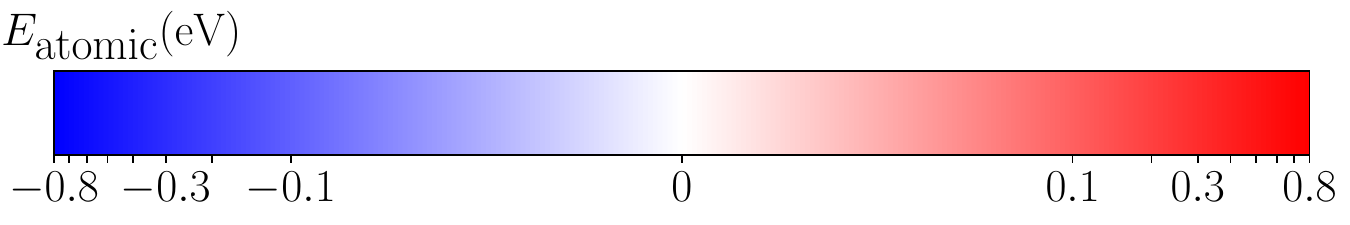}
        \includegraphics[width=\figwidth]{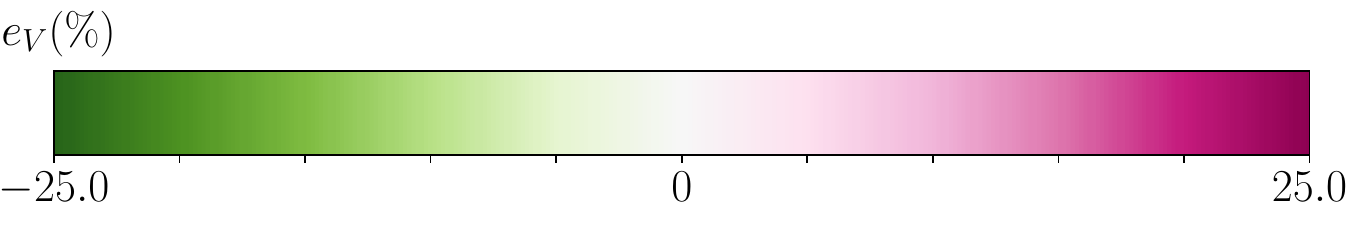}
    }
    \caption{EAM core structures of $a_0[100](010)$ edge, $\frac{a_0}{2}[\bar{1} \bar{1} 1](1\bar{1}0)$ edge, $a_0[100](011)$ edge and $\frac{a_0}{2}[111](1\bar{1}0)$ $71^\circ$ mixed dislocations in BCC Fe. The geometries are relaxed using the EAM potential, with atoms colored based on their atomic energies calculated from the same potential. The black arrows show the differential displacements. The color contours show the distribution of volumetric strains. The ellipses near the dislocation centers mark the widths of the core (listed in \Tab{corewidths}) from the center in and out of the cut planes, calculated using weighted deviation of atom coordinates, with weights defined by the magnitude of difference between EAM and anisotropic elastic atomic energies.}
    \label{fig:core-eam}
\end{figure*}

\begin{figure*}[htbp]
    \makebox[\textwidth][c]{%
        \includegraphics[width=\figwidth]{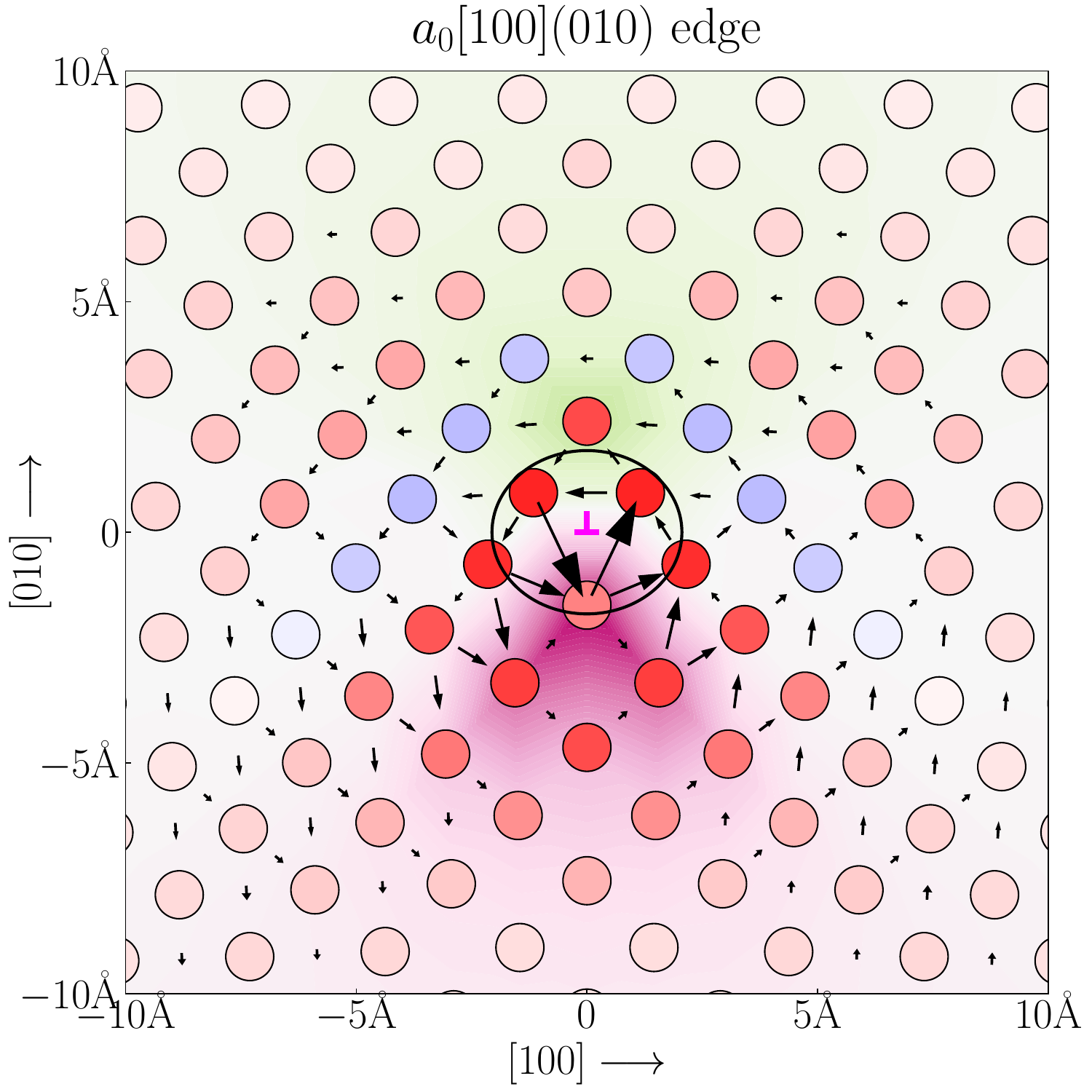}
        \includegraphics[width=\figwidth]{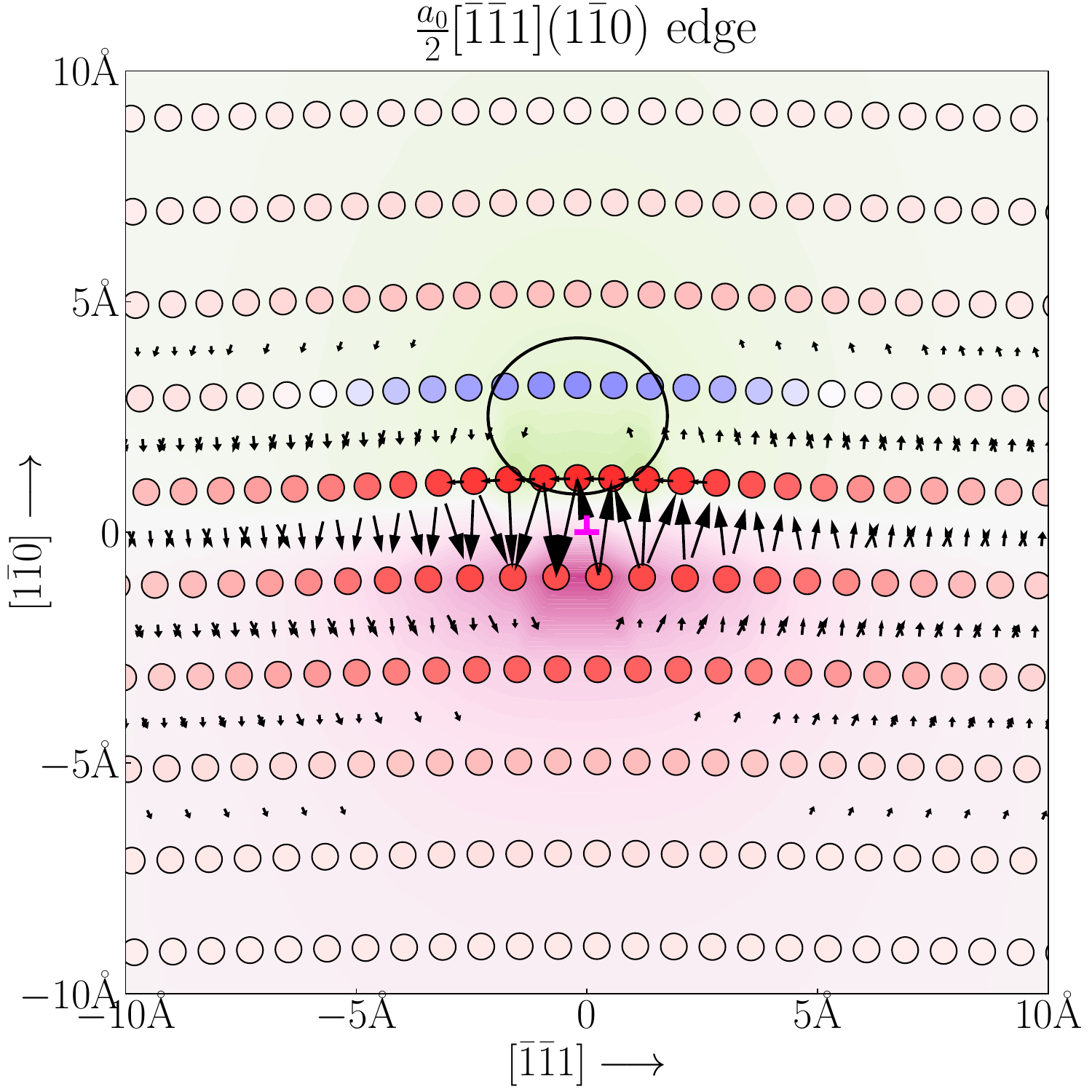}
    }
    \makebox[\textwidth][c]{%
        \includegraphics[width=\figwidth]{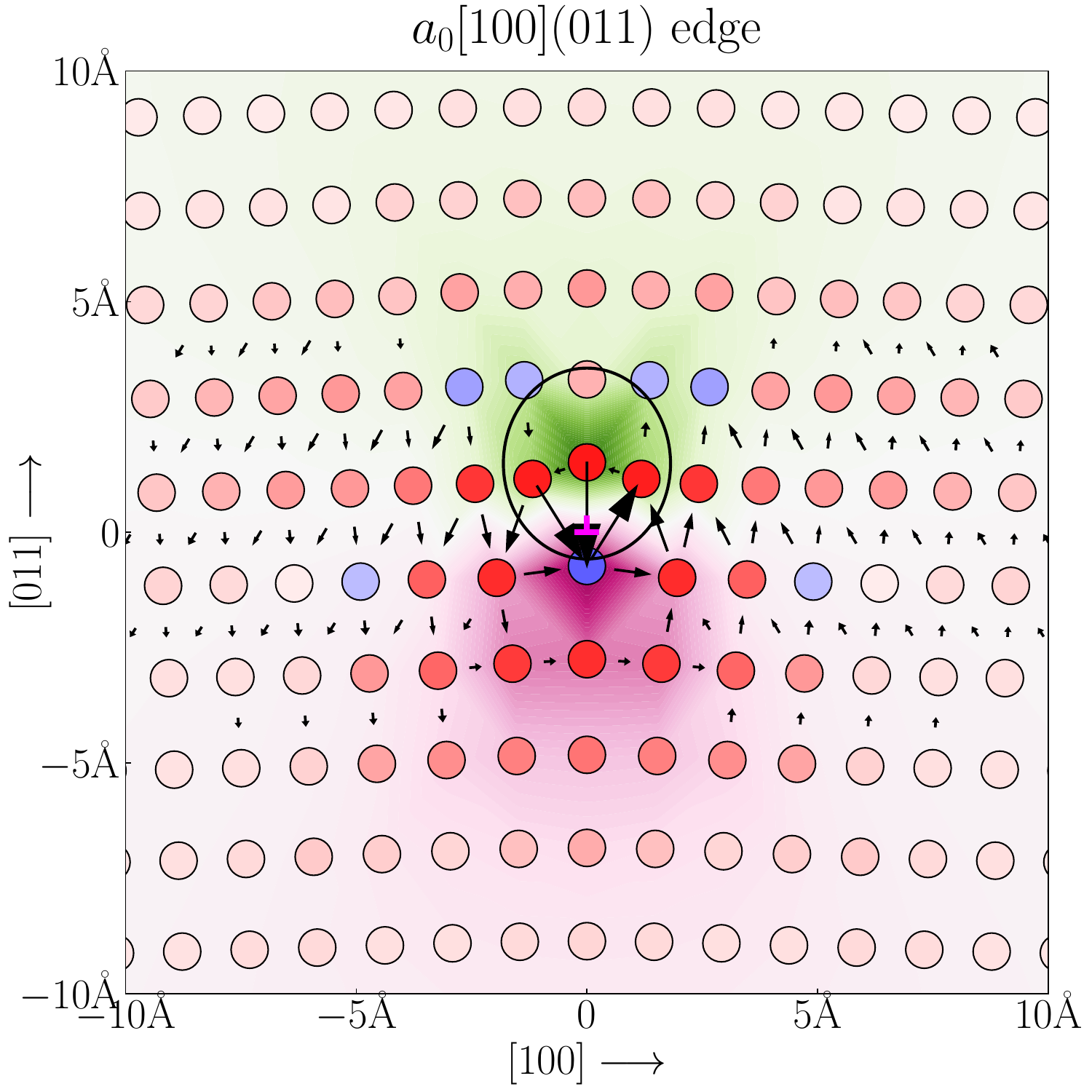}
        \includegraphics[width=\figwidth]{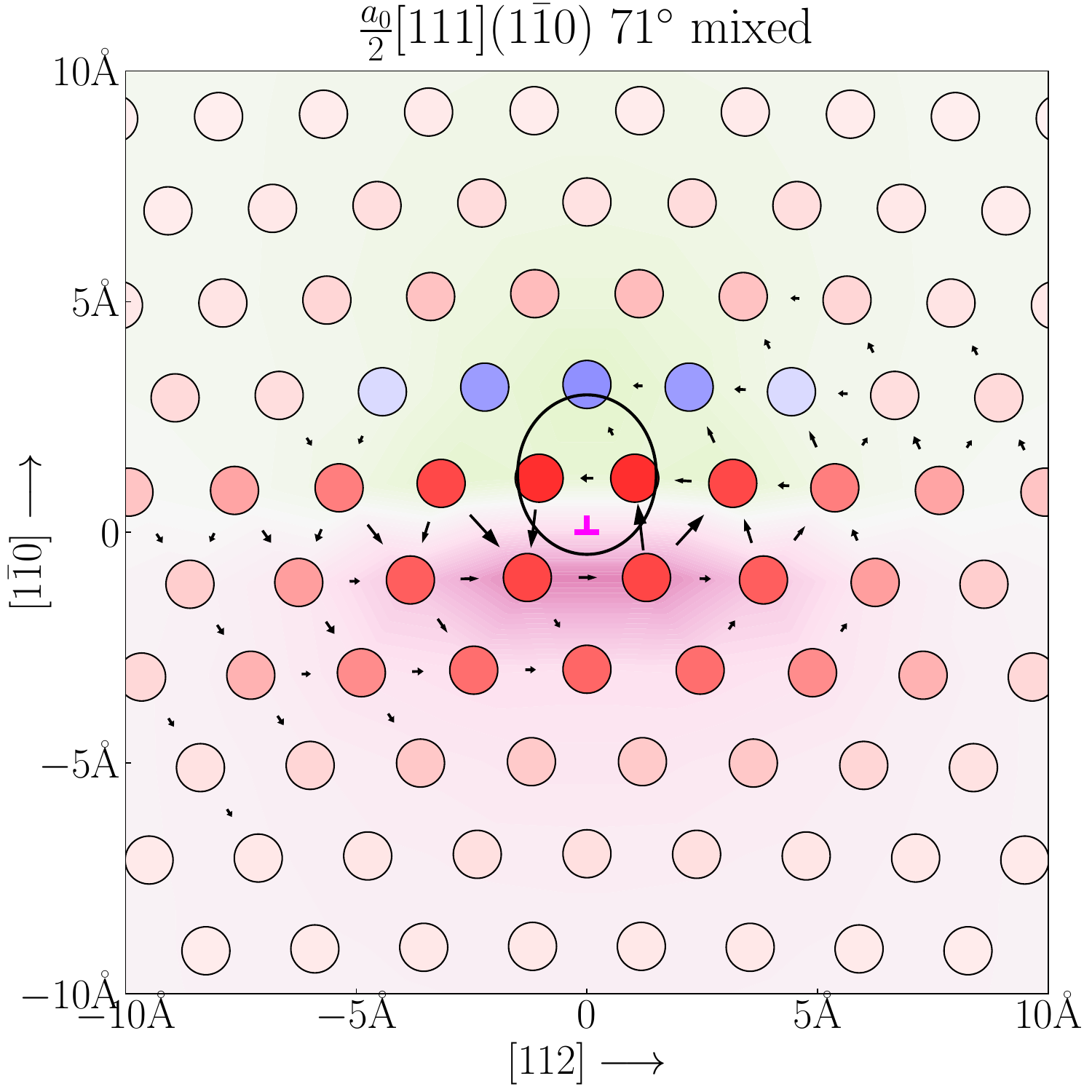}
    }
    \makebox[\textwidth][c]{%
        \qquad
        \includegraphics[width=\figwidth]{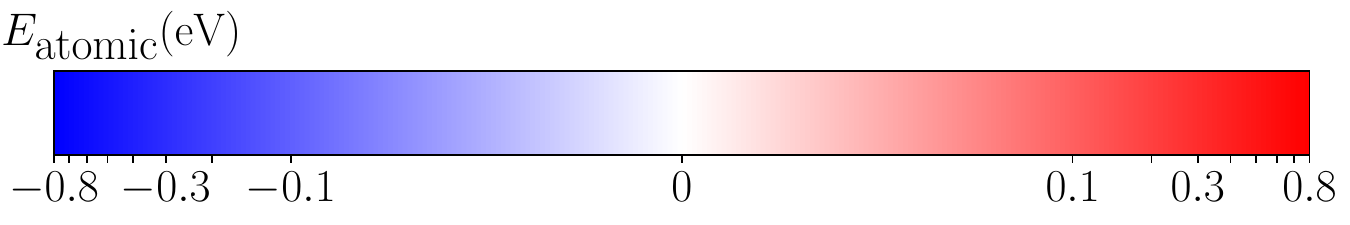}
        \includegraphics[width=\figwidth]{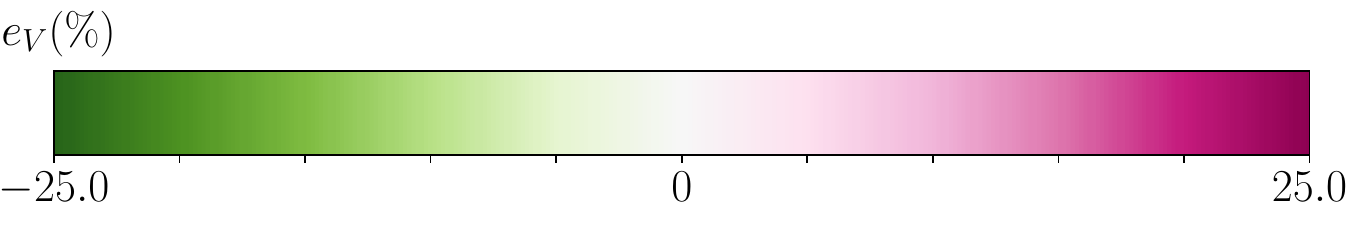}
    }
    \caption{GAP core structures of $a_0[100](010)$ edge, $\frac{a_0}{2}[\bar{1} \bar{1} 1](1\bar{1}0)$ edge, $a_0[100](011)$ edge and $\frac{a_0}{2}[111](1\bar{1}0)$ $71^\circ$ mixed dislocations in BCC Fe. The geometries are relaxed using the GAP potential, with atoms colored based on their atomic energies calculated from the same potential. The black arrows show the differential displacements. The color contours show the distribution of volumetric strains. The ellipses near the dislocation centers mark the widths of the core (listed in \Tab{corewidths}) from the center in and out of the cut planes, calculated using weighted deviation of atom coordinates, with weights defined by the magnitude of difference between GAP and anisotropic elastic atomic energies.}
    \label{fig:core-gap}
\end{figure*}

Relaxations using EAM and GAP potentials lead to similar core structures but different atomic energies compared to the DFT calculations. The EAM and GAP atomic energies differ from the EDM atomic energies by distributing asymmetrically above and below the slip planes, and also by having some negative energies, as revealed in \Fig{core-eam} and \Fig{core-gap}. For both EAM and GAP potentials, the DD maps and volumetric strains have similar distributions compared to the DFT case in each dislocation: the minima and maxima of DDs and volumetric strains emerge near the core center, the DD maps show similar spreading of the core, and the atoms far away from the core center and the slip plane have both small DDs and volumetric strains. This proves that both EAM and GAP potentials predict atom displacements similar to DFT. The atomic energies, however, behave differently from EDM energies in DFT geometries. Core widths are also calculated from  \Eqn{corewidths} and \Eqn{avgpos} with the same weight function using the difference between EAM/GAP per-atom potential energies and anisotropic elastic energies, and visualized as ellipses. For EAM, the $a_0[100](010)$ edge and $a_0[100](011)$ edge dislocations show atomic energies that are smaller in the compressive regime and greater in the tensile regime compared to the corresponding parts in the DFT geometries, causing the centers of the ellipses to be shifted in the negative $n$-direction. In the GAP geometries, negative energies are predicted for several atoms in the compressive regime while the anisotropic elastic energies are positive for all atoms, leading to large relative energy difference $| \Delta E_s / E_s^\text{elas}|$ that shift the centers of the core widths ellipses towards the compressive regime in $a_0[100](011)$ edge, $\frac{a_0}{2}[\bar{1}\bar{1}1](1\bar{1}0)$ edge and $\frac{a_0}{2}[111](1\bar{1}0)$ $71^\circ$ mixed dislocations. In the $a_0[100](010)$ edge dislocation, large $| \Delta E_s / E_s^\text{elas}|$ emerge from both the negative energies in the compressive regime and the large positive energies in the tensile regime, which roughly cancel out and result in the ellipse center to be located near the dislocation center.

\begin{figure*}[htbp]
    \makebox[\textwidth][c]{%
        \includegraphics[width=\wholefigwidth]{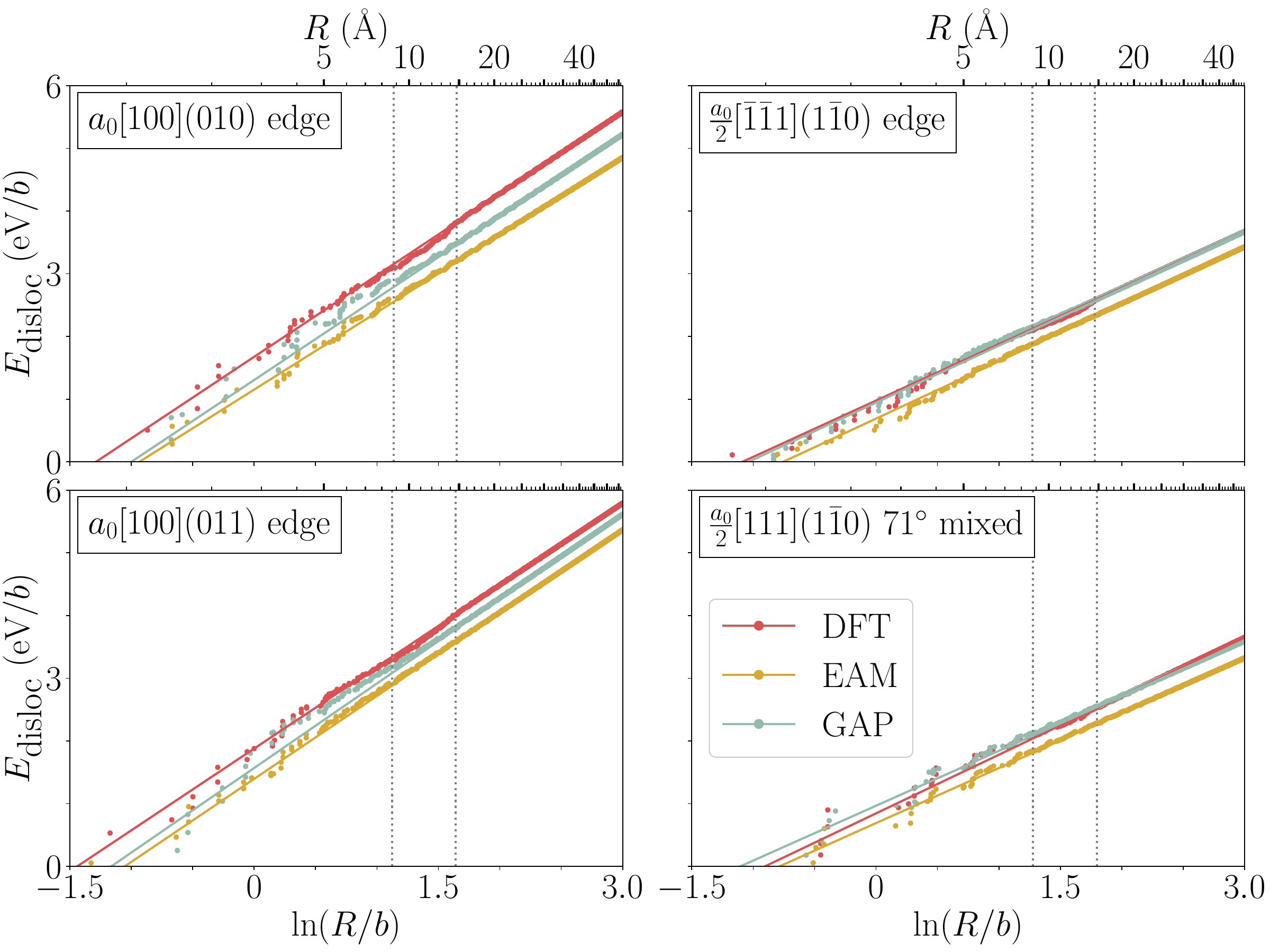}
    }
    \caption{Dislocation line energies of $a_0[100](010)$ edge, $\frac{a_0}{2}[\bar{1} \bar{1} 1](1\bar{1}0)$ edge, $a_0[100](011)$ edge and $\frac{a_0}{2}[111](1\bar{1}0)$ $71^\circ$ mixed dislocations in BCC Fe, calculated using DFT, EAM potential and GAP potential, respectively. The line energies are calculated by integrating the atomic energies within distance $R$ to the dislocation center. For DFT, region I and II atomic energies are calculated using EDM while region III and beyond use anisotropic elastic energies. The vertical dotted lines show the positions of the boundaries that separate regions I/II and II/III. Line energies are fitted and plotted as straight lines with the data points. The linear fittings are done using data points in the range of (2.5, 4.5) for $\ln(R/b)$.}
    \label{fig:lineEs}
\end{figure*}

\Fig{lineEs} shows partial sums of atomic energies to distance $R$ plotted with respect to the logarithmic distance $\ln (R/b)$, showing linear behavior at large $R$. For each DFT geometry, atoms in regions I and II are used, while a slab of radius 300.0 \AA\ with atoms displaced according to the Volterra's equation based on anisotropic elasticity\cite{Bacon1980} is concatenated to the DFT geometry, matching and replacing the atoms in region III and expanding the DFT geometry with atoms beyond region III from the slab. EDM energies are used for atoms in regions I and II while anisotropic elastic energies are used for atoms beyond region II. For EAM and GAP potentials, the per-atom potential energies are used for all atoms up to $r=300.0$ \AA. The line energies $E_\text{disloc}$ are plotted in \Fig{lineEs}, in which linearity is identified for $\ln (r/b)$ falling between 2.5 and 4.5 for all the cases. We do linear fitting to the line energy data based on \Eqn{E_disloc} for the analytical form of the line energy of each dislocation, and the fitted slopes and intercepts are listed in \Tab{lineEs} that correspond to the energy prefactor $\frac{Kb^2}{4 \pi}$ and the core energy $E_\text{core}$. The core energy comes in pair with the choice of the core size $r_c$, and we define the core size as the length of the Burgers vector $r_c = b$ here.

All the three methods predict the same order of dislocations from low to high core energies, with the GAP potential agreeing better with DFT/EDM in numbers compared to the EAM potential, albeit with the discrepancy in the distribution of individual atomic energies. For line energy calculations of the DFT geometries, either EDM or anisotropic elastic energies can be applied to region II atoms, leading to a range of core energies for each geometry, as shown in \Tab{lineEs}. The $a_0[100](010)$ edge and the $a_0[100](011)$ edge dislocations are predicted by all the methods to have higher core energies than the $\frac{a_0}{2}[\bar{1}\bar{1}1](1\bar{1}0)$ edge and the $\frac{a_0}{2}[111](1\bar{1}0)$ $71^\circ$ mixed dislocations. Besides, the core energies of the $\frac{a_0}{2}[\bar{1}\bar{1}1](1\bar{1}0)$ edge and the $\frac{a_0}{2}[111](1\bar{1}0)$ $71^\circ$ mixed dislocations are predicted by all the methods to be very similar, within relative difference of 2.5\%, whereas the core energy of the $a_0[100](011)$ edge dislocation is approximately 20\%\ higher than that of the $a_0[100](010)$ edge dislocation as predicted by the EAM and the GAP potential, and 9\%\ to 16\%\ higher as predicted by EDM with DFT. The core energies predicted by the GAP potential are close to the EDM predictions with DFT in general, while the predictions of the EAM potential are systematically smaller. This is because the large positive atomic energies in the cores of EAM geometries are typically smaller than the corresponding atomic energies in the DFT or the GAP geometries. The energy prefactors are also listed in \Tab{lineEs}, which shows good agreement between the slope of the fitted line energies and the theoretical predictions by anisotropic elasticity. 

We compare our core energies of isolated dislocations with those calculated with dislocation dipoles, showing reasonable comparison with screw dislocation core energies in Clouet's work\cite{Clouet2009Ecore,Clouet2011Ecore} but systematically higher than values reported in Bertin's work\cite{Bertin2021}. DFT calculations of $a_0/2\langle 111 \rangle$-type screw dislocation dipoles show the core energy of $219 \pm 1$ meV/$\text{\AA}$ for an easy core configuration, and $227 \pm 1$ mev/$\text{\AA}$ for a hard core configuration\cite{Clouet2009Ecore}. The \Eqn{reaction} reaction is able to occur when energy of the $a_0 \langle 100 \rangle$ dislocation is lower than the energy of two $a_0/2 \langle 111 \rangle$ dislocations. The core energies predicted by EDM are converted to around 582 meV/\AA\ for the $a_0[100](010)$ dislocation, and 655 meV/\AA\ for the $a_0[100](011)$ dislocation, which are both greater than twice the core energy of either $a_0/2 \langle 111 \rangle$ screw dislocation; however, if we consider the elastic energy contribution $\frac{K_{\langle 111 \rangle}b_{\langle 111 \rangle}^2}{4\pi} \ln \frac{r}{b_{\langle 111 \rangle}}$, and $\frac{K_{\langle 100 \rangle}b_{\langle 100 \rangle}^2}{4\pi} \ln \frac{r}{b_{\langle 100 \rangle}}$ and with the energy prefactors evaluated from anisotropic elasticity in \Tab{lineEs}, one shows that the combined line energies of two $a_0/2 \langle 111 \rangle$ dislocations are greater than the line energy of one $a_0 \langle 100 \rangle$ dislocation once $R$ exceeds a small value of around 5 \AA, which is significantly smaller than the grain sizes in the metal. Therefore, the core energies we obtained support that the reaction $\frac{a_0}{2}[111] + \frac{a_0}{2}[1\bar{1}\bar{1}] \rightarrow a_0 [100]$ is able to occur in practice. Bertin \textit{et al.} calculate the core energies of dislocations of various characters in BCC Fe from dislocation dipoles using classical potentials, including the same Mendelev EAM potential used in this work\cite{Bertin2021}; however their core energies contain an unknown correction term dependent on the dislocation character angle, which nullifies direct comparison with our results which use an isolated dislocation.

\begin{table*}[htbp]
    \caption{The energy prefactor $\frac{Kb^2}{4\pi}$ and core energy $E_\text{core}$ (both in eV/$b$) obtained from the slope and intercept of DFT, EAM and GAP line energies, respectively, as shown in \Fig{lineEs}. Note that DFT, EAM, and GAP have different elastic constants, which leads to different energy prefactors across different methods.}
    \centering
    \begin{tabular}{ccccccccc}
    \hline \hline
    & \multicolumn{2}{c}{DFT} & & \multicolumn{2}{c}{EAM} & & \multicolumn{2}{c}{GAP} \\
    \cline{2-3} \cline{5-6} \cline{8-9}
    & $\frac{Kb^2}{4\pi}$ (fitted, elastic) & $E_\text{core}$ & & $\frac{Kb^2}{4\pi}$ (fitted, elastic) & $E_\text{core}$ & & $\frac{Kb^2}{4\pi}$ (fitted, elastic) & $E_\text{core}$ \\
    \hline
    $a_0[100](010)$ edge & 1.301,1.301 & 1.621--1.676 & & 1.233,1.235 & 1.153 & & 1.306,1.307 & 1.306 \\
    $\frac{a_0}{2}[\bar{1}\bar{1}1](1\bar{1}0)$ edge & 0.899,0.897 & 0.955--0.978 & & 0.913, 0.914 & 0.691 & & 0.909,0.910 & 0.944 \\
    $a_0[100](011)$ edge & 1.305,1.336 & 1.833--1.878 & & 1.324,1.324 & 1.394 & & 1.349,1.351 & 1.568 \\
    $\frac{a_0}{2}[111](1\bar{1}0)$ $71^\circ$ mixed & 0.869,0.864 & 0.968--0.976 & & 0.878,0.879 & 0.690 & & 0.875,0.876 & 0.966 \\
    \hline \hline
    \end{tabular}
    \label{tab:lineEs}
\end{table*}

We apply first-principles energy density method (EDM) with spin polarization to calculate the atomic energies in the geometries of isolated $a_0[100](010)$ edge, $a_0[100](011)$ edge, $\frac{a_0}{2}[\bar{1}\bar{1}1](1\bar{1}0)$ edge and $\frac{a_0}{2}[111](1\bar{1}0)$ $71^\circ$ mixed dislocations in BCC Fe. The geometries are made and optimized using DFT with flexible boundary conditions (FBC)\cite{Fellinger2018}. The atomic energies partial sums give the line energies, whose intercepts give core energies. Our EDM result shows that the $\langle 100 \rangle$-type dislocations, namely $a_0[100](010)$ edge and $a_0[100](011)$ edge dislocations, have relatively high core energies of above 1.6 eV/$b$, whereas the $\langle 111 \rangle$-type $\frac{a_0}{2}[\bar{1}\bar{1}1](1\bar{1}0)$ edge and $\frac{a_0}{2}[111](1\bar{1}0)$ $71^\circ$ mixed dislocations have relatively low core energies of $0.95-0.98$ eV/$b$. The calculations of EDM atomic energies, line energies and core energies do not rely on the type of dislocation or the material, as long as the standard DFT approach applies. Specifically, the estimation of core energy does not require consideration of the elastic ``core field''\cite{Clouet2009Ecore,Clouet2011Ecore} from force dipoles, and the elastic field is calculated fully under the Volterra framework, leading to a potentially more straightforward way of core energy estimation. The core structures predicted by the EAM and the GAP potentials are similar to the DFT predictions as seen from the DD maps and volumetric strains, but the atomic energy distributions differ near the cores. This is partly because the per-atom potential energy is a different way to partition the system energy to individual atoms compared to the EDM approach, and also a possible consequence of the fact that the potentials are fitted without atomic energy data. The EDM atomic energies and core energies can provide additional data for fitting that contribute to the development of new interatomic potentials. The core energies can be used for evaluation of energy barriers in different processes involving dislocation motion, or provide input to simulations of dislocation dynamics, and thus help understanding dislocation behavior. Finally, by tracking the change in EDM atomic and core energies, it is possible to investigate interactions between a dislocation with other defects, such as solutes or other dislocations.

\begin{acknowledgments}
This work is funded by the U.S. National Science Foundation under grant number NSF/MPS-1940303. The research is performed using computational resources provided by the Stampede 2 supercomputer of the Texas Advanced Computing Center (TACC) at The University of Texas at Austin, and by the Illinois Campus Cluster, which is operated by the Illinois Campus Cluster Program (ICCP) in conjunction with the National Center for Supercomputing Applications (NCSA) and supported by funds from the University of Illinois at Urbana-Champaign. The data is available at doi:10.18126/h3jv-yv88 \cite{BCCFe2025Data}.
\end{acknowledgments}


%
\end{document}